\documentclass[10pt,conference]{IEEEtran}
\IEEEoverridecommandlockouts

\usepackage[linesnumbered,ruled,vlined]{algorithm2e}

\usepackage{amssymb}
\usepackage{array}
\usepackage[caption=false,font=footnotesize,labelfont=rm,textfont=rm]{subfig}
\usepackage{stfloats}
\usepackage{url}
\usepackage{verbatim}
\usepackage{graphicx}
\usepackage{ulem}
\usepackage{subscript}
\usepackage{epstopdf}
\usepackage{pifont}
\usepackage{tabularx}
\usepackage{booktabs} 
\usepackage{caption}
\usepackage{multirow}
\usepackage[table]{xcolor}
\usepackage{color}
\usepackage{hyperref}
\usepackage[most]{tcolorbox}
\usepackage{listings}
\usepackage{xcolor}
\usepackage{subfig}
\usepackage[dvipsnames]{xcolor}
\usepackage{listings}
\usepackage{minted}
\usepackage{makecell}
\usepackage{caption} 
\definecolor{tablegray}{gray}{0.9}
\def\BibTeX{{\rm B\kern-.05em{\sc i\kern-.025em b}\kern-.08em
    T\kern-.1667em\lower.7ex\hbox{E}\kern-.125emX}}
\begin{document}

\newcommand{\highlight}[2][lightgray]{%
    \begin{tabular}{l}
        \rowcolor{#1}%
        #2%
    \end{tabular}
}

\definecolor{mygreen}{rgb}{0, 0.6, 0} 

\newcommand{\mycomment}[1]{%
    \textcolor{mygreen}{\footnotesize // #1}%
}

\SetKwInput{KwInput}{Input}{}{}
\SetKwInput{KwOutput}{Output}{}{}
\title{RSFuzz: A Robustness-Guided Swarm Fuzzing Framework Based on Behavioral Constraints}
\author{\IEEEauthorblockN{Ruoyu Zhou$^{1,2}$,
                          Zhiwei Zhang$^{*1,2}$,                  
                          Haocheng Han$^{1,2}$,
                          Xiaodong Zhang$^{*3}$,
                          Zehan Chen$^{1,2}$,\\
                          Jun Sun$^{4}$,
                          Yulong Shen$^{1,2}$,
                          and Dehai Xu$^{5}$}
\IEEEauthorblockA{$^1$ School of Computer Science and Technology, Xidian University, Xi'an, China}
\IEEEauthorblockA{$^2$ Shaanxi Key Laboratory of Network and System Security, Xidian University, Xi'an, China}
\IEEEauthorblockA{$^3$ University of Science and Technology of China, China}
\IEEEauthorblockA{$^4$ Singapore Management University, Singapore}
\IEEEauthorblockA{$^5$ Yiqiyin (Hangzhou) Technology Co., Ltd. Xi'an Branch, Xi'an, China}
\IEEEauthorblockA{\{ruoyuzhou, xd\_hchan, zehanchen\}@stu.xidian.edu.cn,\\
\{zwzhang, ylshen\}@xidian.edu.cn, zhangxiaodong@ustc.edu.cn, junsun@smu.edu.sg, 15504446630@163.com}
\thanks{* Corresponding Authors.}
}

\maketitle

\begin{abstract}
Multi-robot swarms play an essential role in complex missions including battlefield reconnaissance, agricultural pest monitoring, as well as disaster search and rescue. Unfortunately, given the complexity of swarm algorithms, logical vulnerabilities are inevitable and often lead to severe safety and security consequences. Although various methods have been presented for detecting logical vulnerabilities through software testing, when they are used in swarm environments, these techniques face significant challenges: 1) Due to the swarm's vast composable parameter space, it is extremely difficult to generate failure-triggering scenarios, which is crucial to effectively expose logical vulnerabilities; 2) Because of the swarm's high flexibility and dynamism, it is challenging to model and evaluate the global swarm state, particularly in terms of cooperative behaviors, which makes it difficult to detect logical vulnerabilities.

In this work, we propose RSFuzz, a robustness-guided swarm fuzzing framework designed to detect logical vulnerabilities in multi-robot systems. It leverages the robustness of behavioral constraints to quantitatively evaluate the swarm state and guide the generation of failure-triggering scenarios. In addition, RSFuzz identifies and targets key swarm nodes for perturbations, effectively reducing the input space. Upon the RSFuzz framework, we construct two swarm fuzzing schemes, Single Attacker Fuzzing (SA-Fuzzing) and Multiple Attacker Fuzzing (MA-Fuzzing), which employ single and multiple attackers, respectively, during fuzzing to disturb swarm mission execution. We evaluated RSFuzz's performance with three popular swarm algorithms in simulated environments. The results show that RSFuzz outperforms the state-of-the-art with an average improvement of 17.75\% in effectiveness and a 38.4\% increase in efficiency. We also validated some detected vulnerabilities in real-world environments. Our code and data are publicly available.
\end{abstract}


\section{Introduction}
Multi-robot swarms, consisting of several collaborating robots, are often deployed to perform complex tasks that would be difficult or impossible for a single robot to accomplish \cite{cheraghi2022past, dorigo2013swarmanoid,madridano2021trajectory,gautam2012review}. Currently, swarms are used in diverse applications such as military operations, agriculture, and disaster relief \cite{chung2018survey}. 

A swarm algorithm orchestrates the behaviors of robots of the swarm, dynamically modulating their actions to meet diverse internal and external objectives. For example, in a swarm, each unit is calibrated to reach its destination, avoid obstacles, and maintain safe inter-drone distances. However, vulnerabilities in the control logic may hinder the achieving of these objectives, when exploited—intentionally or otherwise—can lead to system failures or mission compromise \cite{jung2021swarmbug,jung2022swarmflawfinder,deng2021investigation}. In critical operations, such deficiencies often cause significant human or financial losses \cite{mekruksavanich2018applied}. For instance, the Australian Transport Safety Bureau reported a 38\% increase in drone control loss incidents between 2018 and 2019 \cite{ghasri2021factors}. By the end of 2023, the number of registered drones in China surpassed 1.2 million, with a cumulative flight time of approximately 23 million hours \cite{CAAC2023}. These statistics highlight the critical need for comprehensive testing of swarm algorithms to ensure both operational safety and mission success.

Testing multi-robot swarm algorithms faces two major challenges due to their inherent complexity. First, the vast composable parameter space of the swarm makes it extremely difficult to generate failure-triggering scenarios, which are essential for effectively exposing logical vulnerabilities. Second, due to the swarm's high flexibility and dynamism, accurately modeling and evaluating the global swarm state—particularly cooperative behaviors among robots—is challenging. This complexity complicates the detection of logical vulnerabilities.

Early attempts usually employed simple methods like taint tracking and trial-and-error to detect logical vulnerabilities \cite{yuan2014cloudtaint, kreindl2019towards}. However, due to complex parameter dependencies and high dynamism of swarm algorithms, these methods are time-consuming and ineffective. Some researchers have pursued automated testing methods. For example, Jung et al. \cite{jung2021swarmbug,jung2022swarmflawfinder} proposed a fuzzing method based on the Degree of Causal Contribution (DCC) to identify logical vulnerabilities in swarms, although their approach still suffers from low precision and high computational complexity. Deng et al. \cite{deng2021investigation} introduced a Signal Temporal Logic (STL)–based fuzzing method for detecting Byzantine threats. Although it's effective, its real-world implementation remains challenging. Moreover, model-based testing methods, such as those proposed by Kim et al. \cite{kim2021pgfuzz} for tracing control semantic errors, are useful for post-incident analysis but lack real-time applicability for testing swarm logical vulnerabilities. Besides, existing fuzzing techniques primarily target binary vulnerabilities or input validation bugs \cite{kim2019rvfuzzer,feng2020p2im,schiller2023drone}, but they fall short in detecting logic vulnerabilities unique to swarm behavior.

To address the challenge of detecting logical vulnerabilities in multi-robot swarm systems, we propose RSFuzz, a robustness-guided fuzzing framework. RSFuzz leverages attack drones to perform non-contact interference during mission execution, enabling the exposure of latent logical flaws in swarm algorithms. Unlike prior approaches, RSFuzz introduces swarm robustness, a novel metric based on behavioral constraints, to quantitatively assess the system state and guide fuzzing input generation in real time. Furthermore, we utilize the \textit{Katz} method \cite{nathan2017dynamic} to identify key nodes within the swarm, effectively narrowing the input search space and improving testing efficiency. RSFuzz offers four key advantages. \textbf{1) Precise Guidance:} STL-based swarm robustness accurately captures the mission to guide attack drone positioning. \textbf{2) Low Overhead:} By focusing on the swarm's global state rather than considering individual drone behaviors, RSFuzz reduces the computational cost and avoids exhaustive per-drone reasoning. \textbf{3) Optimized input space searching:} Identifying key nodes significantly narrows the test case search space. \textbf{4) Scalability:} RSFuzz is suitable for various scenarios ranging from small-scale operations to large-scale search and rescue missions.
Building upon RSFuzz, we implemented two fuzzing schemes: SA-Fuzzing and MA-Fuzzing. SA-Fuzzing focuses on a single attack drone to efficiently explore failure-triggering scenarios with low computational overhead. MA-Fuzzing, despite involving multiple attack drones in total, activates only one attacker per iteration to maintain focused testing while benefiting from a broader set of candidate attack positions generated across the swarm's key nodes. This approach allows MA-Fuzzing to introduce more diverse and globally-informed perturbations, enhancing its ability to uncover subtle or complex logical vulnerabilities. We evaluated both schemes on three swarm control algorithms \cite{agishev2019adaptive,carnelli2017SwarmRoboticsSim,howard2020swarm} in simulation and validated vulnerabilities in real-world tests using Bitcraze Crazyflie 2.1+ drones \cite{crazyflie}.

Our main contributions are summarized as follows:
\begin{itemize}
    \item[$\bullet$]We propose a novel robustness-guided fuzzing framework based on behavioral constraints. As far as we know, we are the first to leverage the robustness of behavioral constraints to quantify the swarm state, which can optimize the positioning of the attack drone. 
    \item[$\bullet$]We introduce a key node identification method based on \textit{Katz} to narrow the search space for positioning the attack drone. To accommodate diverse fuzzing scenarios, we design two fuzzing schemes: SA-Fuzzing and MA-Fuzzing.
    \item[$\bullet$]We implement SA-Fuzzing scheme and MA-Fuzzing scheme, and have conducted extensive and comparative evaluations on three swarm algorithms on their corresponding simulations and validated the vulnerabilities in real-world environments. All the source code and data are available at \url{https://github.com/Ruoyyy/RSFuzz}. 
\end{itemize}


\section{Background, Threat Model and Motivation}

\subsection{Background and Scope}
\textbf{Swarm Algorithms and Swarm Logical Vulnerabilities.}
A swarm algorithm coordinates multiple drones to conduct the mission's objectives. In this paper, we consider swarm algorithms that manage logic for both individual drones and the swarm's cooperative behaviors. A logic vulnerability in swarm algorithms refers to inherent errors or weaknesses in the design or implementation that can lead to unintended or suboptimal behavior under specific conditions \cite{jung2022swarmflawfinder}. Such logic vulnerabilities compromise the efficiency and reliability of the algorithm, potentially undermining its intended functionality. It is critical to identify these vulnerabilities to ensure the algorithm's effectiveness in real-world applications.

\subsection{Threat Model}
We consider a threat scenario in which an adversary controls one or more attack drones capable of approaching the target swarm. The adversary is assumed to have knowledge of the swarm's mission and control algorithm but does not have access to the software or hardware of the target drones. Instead of launching traditional attacks such as GPS jamming~\cite{kerns2014unmanned,seo2015effect}, firmware tampering~\cite{kim2021pgfuzz}, or software exploits~\cite{kim2019rvfuzzer}, the adversary leverages subtle, non-contact behavioral interference—positioning attack drones to influence swarm behavior without causing direct collisions.

Such threat scenarios are increasingly feasible due to the low cost and high availability of commercial drones, which adversaries can easily repurpose for stealthy and repeatable operations. Compared to hardware- or communication-level attacks, logic-level behavioral interference is both more covert and cost-effective, making it an appealing vector for adversaries in both civilian and military contexts. 

This work focuses specifically on distributed, autonomously controlled drone swarms. These systems are widely deployed and represent high-value multi-robot platforms. Their highly coordinated and dynamic behaviors amplify the impact of even minor logical vulnerabilities, potentially resulting in large-scale mission failure or severe safety and operational risks.


\subsection{Motivation Example}

\begin{figure}[htpb]
    \centering
    \includegraphics[width=\linewidth]{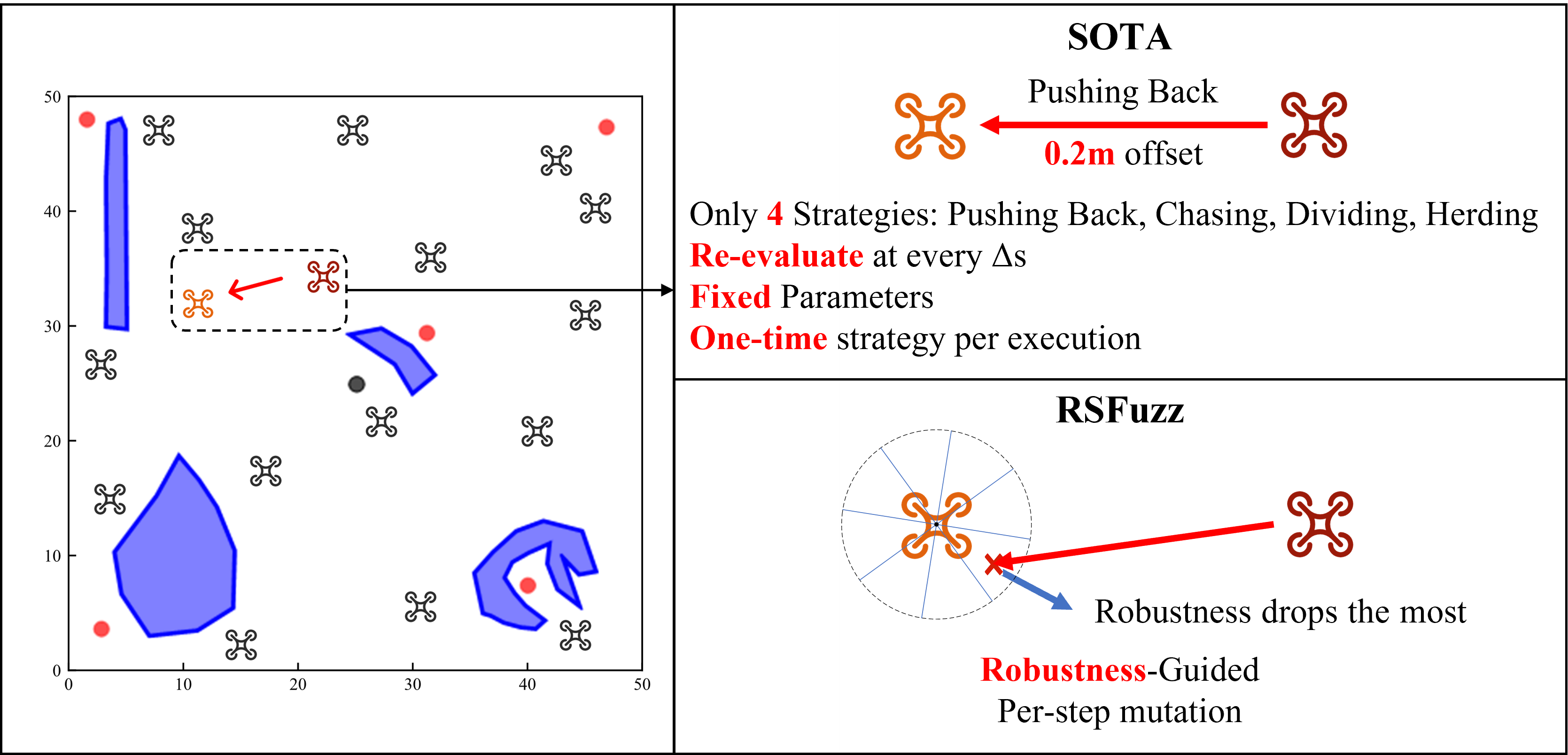}
    \caption{Motivation illustration}
    \label{fig_1}
\end{figure}
We use an example to show how RSFuzz works and its advantages. As shown in Fig. \ref{fig_1}, there is a swarm including 15 drones, its mission is to locate the earthquake victims (represented as five red points) within a certain time limit. We use an attack drone (i.e., the red drone)to interfere with the execution of the swarm so that some drones in the swarm collide with each other, crash into obstacles, or fail to complete the mission within the specified time.


\textbf{Limitations of SOTA:} SWARMFLAWFINDER \cite{jung2022swarmflawfinder} is the state-of-the-art swarm fuzzing tool based on DCC to guide attack strategies, and yet it suffers from two issues.
\begin{itemize}
    \item Scalability issues: DCC's computational complexity grows exponentially with the swarm size and mission duration. In our example, with 15 drones executing a 700-second mission comprising 1,400 iterations, each test case requires the execution of 1,400 × 19 auxiliary experiments, accounting for 15 additional swarm drones and 4 environmental factors, making causal inference prohibitively time-consuming. This inefficiency is exacerbated in large-scale scenarios with extended mission durations.
    \item Limited attack surface: The method relies on only four fixed attack strategies (e.g., pushing back, chasing), with preset parameters (e.g., attack distances 0.2m offset), lacking adaptability to the evolving state of the swarm. Moreover, each execution uses a single attack strategy throughout the entire mission, and the mutation of strategies only occurs between executions rather than within them, limiting its ability to dynamically adjust to swarm behaviors during runtime.
\end{itemize}

\section{Design of RSFuzz}

\begin{figure*}[htpb]
    \centering
    \includegraphics[width=\textwidth]{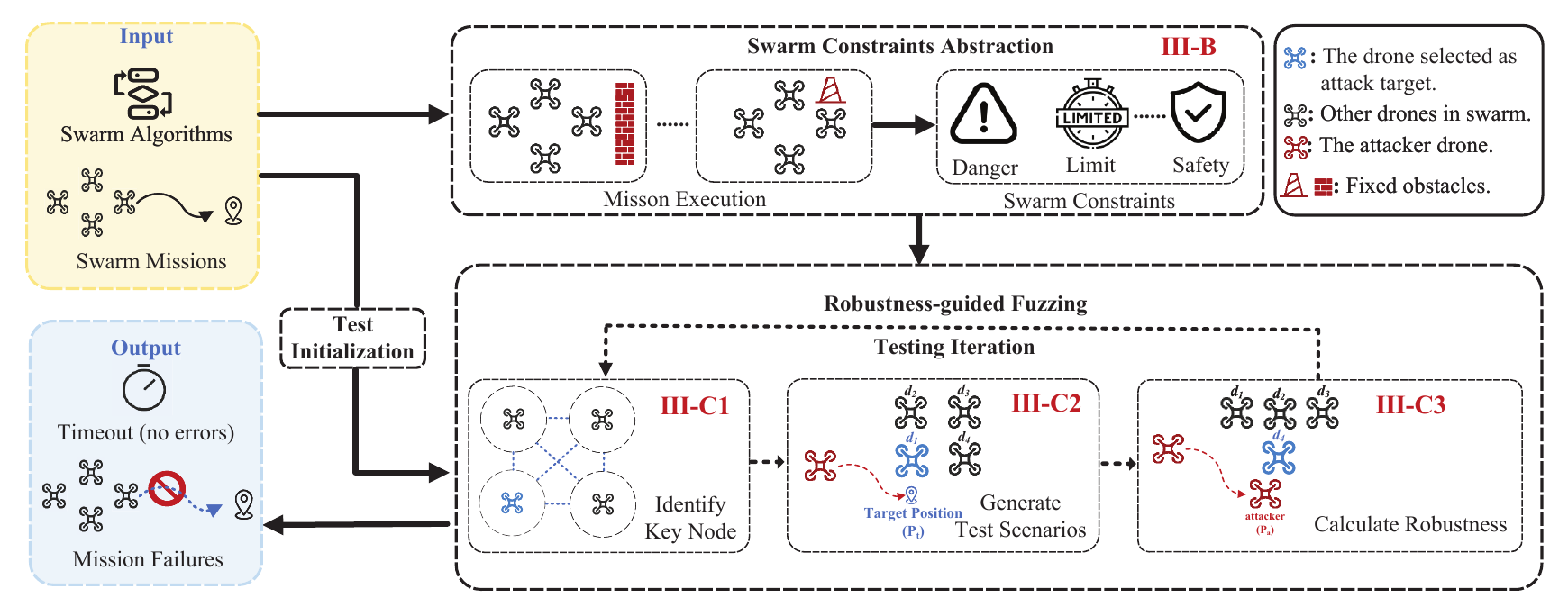}
    \caption{Overview of RSFuzz}
    \label{41}
\end{figure*}
\subsection{RSFuzz Framework}
The overview of RSFuzz is shown in Fig. \ref{41}. The input is the swarm algorithm and mission configuration, while the output is a set of logical vulnerabilities or a timeout (if no errors are found). The gray shaded area represents the main components of the fuzzer. First, the swarm's STL constraints are manually extracted from normal mission execution (III-B). Second, in order to narrow the space of test cases generation, RSFuzz identifies the key node in the swarm based on the \textit{Katz} method \cite{truszkowski2004survey,winfield2005formal} (III-C1). Then, it introduces attack drones to target these key nodes and generate test cases that lead to failures (III-C2). Finally, it calculates the swarm's robustness through constraints, thereby uncovering logical vulnerabilities. Additionally, RSFuzz mutates the generated tests to create new test cases and repeats this until timeout (III-C3).

\subsection{Swarm Constraints Abstraction}
During the execution of missions, swarms must satisfy various constraints. These constraints can be categorized into safety constraints and mission constraints. Safety constraints ensure the swarm does not experience collisions or crashes. Mission constraints ensure the swarm achieves its missions. Drawing from an extensive literature review and practical scenario analysis, we abstract five representative constraints that form the basis of our formal specifications (Equation~\ref{eq:stl_rules}).
\begin{equation}
\footnotesize
    \left\{
    \begin{aligned}
    \varphi_1 &\equiv \square\left( d(t) \geq d_s \right) \\
    \varphi_2 &\equiv \square\left( 0 \leq v(t) \leq v_s \right) \\
    \varphi_3 &\equiv \square\left( |a(t)| \leq a_s \right) \\
    \varphi_4 &\equiv \square\left( d_m \leq \| x(t) - x_{\text{near}}(t) \|_2 \leq d_M \right) \\
    \varphi_5 &\equiv \square\left( \|x_{avg}(t + \Delta t )-x_{\text {g}}\|_2 <  \|x_{avg}(t)-x_{\text {g}}\|_2 \right)
    \end{aligned}
    \right.
    \label{eq:stl_rules}
\end{equation}

As shown in Equation~\ref{eq:stl_rules}, the set of formulas $\varphi_1$ through $\varphi_5$ encode key behavioral constraints using STL. In this context, the operator $\square$ denotes the “always” temporal modality, meaning that the constraint must hold at every time step throughout the mission.

\begin{itemize}
    \item \textbf{Distance Constraint ($\varphi_1$).} Ensures that each drone maintains a minimum safe distance $d_s$ from surrounding obstacles at all times~\cite{kumar2021velocity,Fan2022Adversarial}. Here, $d(t)$ denotes the real-time distance between a drone and an obstacle at time $t$, and $d_s$ is the threshold for safe separation.
    
    \item \textbf{Velocity Constraint ($\varphi_2$).} Requires that the velocity of each drone remain below the predefined safety limit $v_s$~\cite{kumar2021velocity,raghuwaiya2018leader,zhang2013ucav}. Specifically, $v(t)$ denotes the velocity of a drone at time $t$, and $v_s$ is the maximum allowed speed.
    
    \item \textbf{Acceleration Constraint ($\varphi_3$).} Limits the absolute acceleration $a(t)$ of each drone to a safe threshold $a_s$ to ensure smooth and controlled motion~\cite{he2018feedback}.
    
    \item \textbf{Separation and Cohesion constrain ($\varphi_4$).} Maintains swarm cohesion by requiring the Euclidean distance between a drone and its neighboring agents to remain within the bounds $[d_m, d_M]$. Here, $x(t)$ denotes the position of a drone, and $x_{near}(t)$ refers to the position of a nearby swarm member~\cite{arafat2019localization}.
    
    \item \textbf{Goal-Convergence Constraint ($\varphi_5$).} This constraint ensures that the swarm consistently moves toward the mission target $x_g$~\cite{ducatelle2011self}. The average drone position over each interval $[t, t+\Delta t]$ is defined as:
    \[
    x_{avg}(t) = \left\| \operatorname{avg}(x(t : t+\Delta t)) \right\|
    \]
    where $x(t : t+\Delta t)$ denotes the positions of the drone during the interval. The constraint requires that the swarm's average position in the next interval is closer to the goal than in the current one. This encourages continuous progress toward the goal across time.
\end{itemize}
 
\subsection{Robustness-guided Fuzzing}
Both schemes follow a structured process that includes test definition (Input and Output), test initialization, test execution, and test iteration.

\textbf{Test Initialization:} RSFuzz begins by setting up the tested swarm and mission scenario according to the swarm algorithm and mission configuration. It then identifies the key node within the swarm as the target drone $P_t$. The attack drone is initially positioned outside the swarm's sensing range to ensure a more realistic scenario and maintain the test's validity. 

To accurately assess the degree to which a swarm adheres to behavioral constraints, we propose a metric called \textbf{swarm robustness}, which is computed based on the satisfaction of temporal logic constraints. Furthermore, we quantitatively evaluate the changes in constraint satisfaction over time based on STL robustness semantics proposed in RTAMT \cite{nivckovic2020rtamt}. RSFuzz uses swarm robustness as feedback to guide the fuzzing, enabling it to dynamically identify critical states that are close to violating constraints. The specific calculation of swarm robustness is shown in Equation \ref{equ2}.

First, RSFuzz calculates the degree to which each drone complies with behavioral constraints. In the Equation \ref{equ2}, $rob_1$ to $rob_5$ correspond to five constraints. Then, for each drone, RSFuzz normalizes and sums the robustness across the five dimensions to obtain the individual robustness $R_i$ (where $i$ is the index of drones in the swarm). The calculation of $R_i$ is detailed in Equation \ref{equ3}. Finally, swarm 
robustness $\mathbb{R}$ is derived by aggregating the individual robustness of all drones in the swarm. The calculation of $\mathbb{R}$ is detailed in Equation \ref{equ4}.

\begin{equation}
    \footnotesize
    \left\{     
	\begin{aligned}
	rob_1 &= d(t)-d_s\\
        rob_2 &= \min \left(v(t), v_s-v(t)\right)\\
        rob_3 &= \min \left(|a(t)|-a_s, 0\right)\\
        rob_4 &= \max \left(\left\|x(t)-x_{\text {near }}(t)\right\|_2-d_m, d_M-\left\|x(t)-x_{\text {near }}(t)\right\|_2\right)\\
        rob_5 &= \left\|x_{\text {avg }}(t+\Delta t)-x_{\text {des }}(t)\right\|_2-\left\|x_{\text {avg }}(t)-x_{\text {des }}(t)\right\|_2
	\end{aligned}
    \right.
    \label{equ2}
\end{equation}

\begin{equation}
\footnotesize
    R_j=\sum_{i=1}^{5}{\frac{rob_i}{Max(rob_1, rob_2,\dotsb, rob_5)}}
    \label{equ3}
\end{equation}

\begin{equation}
\footnotesize
    \mathbb{R}=\frac{\sum_{j=1}^{n} R_j}{n}
    \label{equ4}
\end{equation}

\subsubsection{Identifying Swarm Key Node}
In a swarm, certain individuals have a much greater impact on swarm robustness than others. The individual with the most significant influence is referred to as the swarm key node. Focusing on this key node improves the efficiency of fuzzing. To efficiently identify the swarm key node, we first need a model that captures how influence propagates. Our model is built on the principle that swarm influence is spatially localized. Therefore, instead of a computationally expensive fully-connected graph, we construct a sparse swarm constraint influence graph, $G = (V, E, W)$.
In this graph, an edge $(d_i, d_j)$ exists if the Euclidean distance $\|p_i - p_j\|_2$, is below an adaptive influence threshold $\tau$. The weight $w_{ij}$ of an existing edge is the inverse of this distance:
\begin{equation} \label{eq:weight}
w_{ij} = \frac{1}{\|p_i - p_j\|_2}
\end{equation}

We define $\tau$ based on the swarm's nominal operational distance, which is typically determined by the algorithm's communication range ($D_{comm}$) or the mission's pre-defined formation separation distance ($D_{form}$). Specifically, we set:
\begin{equation} \label{eq:threshold}
\tau = 1.5 \cdot D_{nominal}
\end{equation}
where $D_{nominal}$ can be either $D_{comm}$ or $D_{form}$, depending on the algorithm's context. The details of $D_{nominal}$ are explained in \nameref{Evaluation}.

\begin{figure}[htpb]
    \centering
    \includegraphics[width=\linewidth]{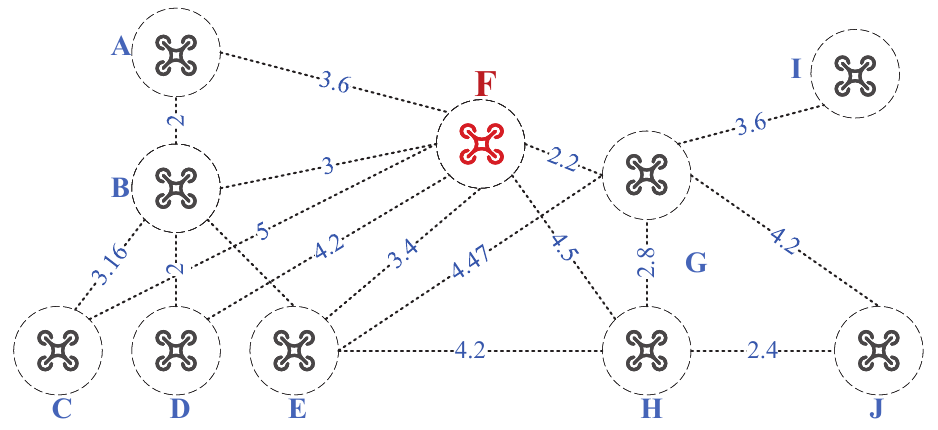}
    \caption{Example of swarm key node} 
    \label{32} 
\end{figure}

RSFuzz then applies \textit{Katz} centrality analysis on this graph to identify the most influential node. The centrality score $x_i$ for each node $i$ is calculated iteratively based on the contributions of its neighbors, formally defined as:
\begin{equation}
    \label{eq:katz_centrality}
    x_i = \alpha \sum_{j=1}^{N} w_{ji} x_j + \beta
\end{equation}
where $\alpha$ is an attenuation factor that controls the influence of distant neighbors, $\beta$ is a constant bias term to provide each node with a base level of importance, and $w_{ji}$ represents the weight of the edge from node $j$ to node $i$. The node with the highest centrality score $x_i$ is then designated as the swarm key node~\cite{bloch2023centrality,zhan2017identification}. This node is expected to have a disproportionately large effect on constraint satisfaction or violation under perturbation. Fig.~\ref{32} shows a conceptual example to illustrate how the key node \textit{F} is identified through centrality analysis.

\subsubsection{Generate Test Scenarios}
In RSFuzz, each fuzzing input is defined as a test scenario represented by a tuple \textless \emph{$P_t$, $P_a$}\textgreater. The first step is to initialize the test according to the definition. The specific definitions of $P_t$ and $P_a$, as well as the method of test initialization, are described as follows.
\begin{itemize}
\item Target Drone Position ($P_t$): The position of the drone targeted by the attack. Clearly, if $P_t$ has the greatest impact on swarm robustness, the input space for testing can be significantly reduced, enhancing testing efficiency. 

\item Attack Drone Initial Position ($P_i$): The initial position vector of the attack drone, denoted by $P_i \in \mathbb{R}^n$, where the dimension $n=2$ or $n=3$. The position is initialized outside the swarm's designated safety range to prevent immediate collisions.

\item Attack Drone Position ($P_a$): The position vector of the attack drone, denoted by $P_a \in \mathbb{R}^n$, where the dimension $n=2$ or $n=3$. The domain of $P_a$ excludes the swarm's safety range.
\end{itemize}
\begin{figure}[htpb]
    \centering
    \subfloat[Drone's sensing area]{\label{42a}\includegraphics[width=0.2\textwidth]{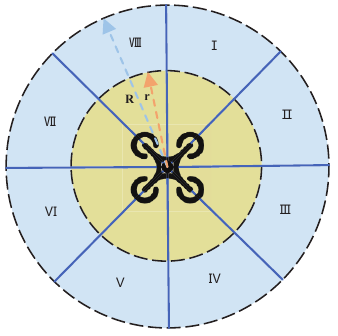}}
    \subfloat[Attacker alternate generation areas]{\label{42b}\includegraphics[width=0.25\textwidth]{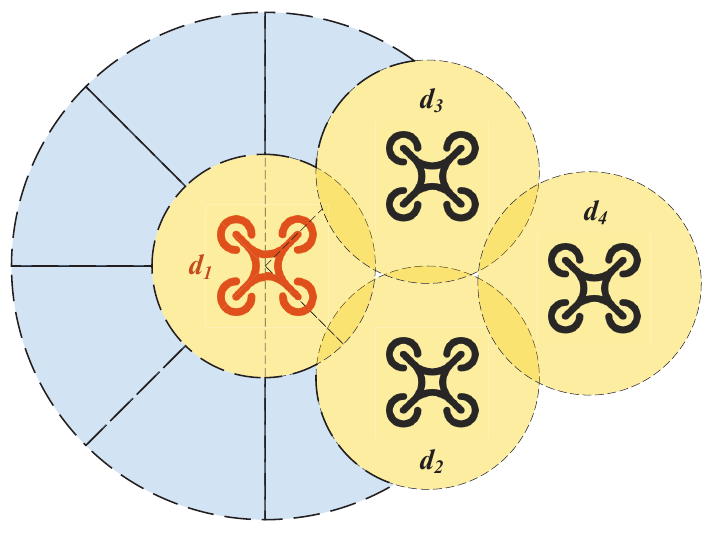}}
    \caption{Drone sensing range and attack deployment positions.}
    \label{42}
\end{figure}

\subsubsection{Execute Test and Calculate Robustness}

An example of the attack drone's generation is shown in Fig. \ref{42}. As shown in Fig. \ref{42a}, the inner circle represents the sensing area of the drone with a radius $r$, while the outer circle marks the maximum distance between the attacker and the target with a radius $R$. Consequently, the attack drone's generation area lies between the inner and outer circles, spanning a radius of $R-r$, and this space is equally divided into $n$ partitions. It is considered equivalent to generate the attack drone at any point within each partition. Note that the attack drone must maintain a basic safety distance from other drones, so any regions that fail to meet this constraint are excluded. In Fig. \ref{42b}, the swarm consists of drones $d_1$ to $d_4$, with $a$ chosen as the target drone. Due to safety distance constraints (highlighted in yellow), the attack drone can only be generated in areas where there is no conflict between the sensing range and the safety distance (marked in blue). RSFuzz then selects the position that results in the greatest reduction in swarm robustness as the initial $P_t$ and sets the attack drone's initial position $P_i$ slightly beyond the swarm's sensing range, at a location farther from $P_t$. This completes the test initialization.

Given the current position $P_a$ of the attack drone and its maximum velocity $v$, SA-Fuzzing first determines the attack drone's reachable area. The sensing area of each drone in the swarm is defined by a fixed detection radius. SA-Fuzzing then selects a subset of drones $D_{vx}$ whose sensing areas intersect with the attack drone's reachable area to construct a constraint influence subgraph. The key node of this subgraph, identified using \textit{Katz} centrality, is selected as the target drone $P_t$.

Next, SA-Fuzzing randomly samples $n$ candidate positions for the attack drone from the intersection of $P_t$'s sensing area and the attack drone's reachable area. For each candidate position, it simulates the swarm's next state and calculates the corresponding robustness value. If any drone is killed during this execution, all the test cases from the current fuzzing execution are immediately output. Otherwise, the candidate position with the minimum robustness score is selected as the next $P_a$, and the fuzzing process continues.

\textbf{Testing Iteration. }The test execution is repeated until the swarm mission either fails or completes successfully. If a failure occurs, the test case causing the error is outputted; otherwise, if the mission completes without failure, it indicates that no logical vulnerabilities were detected.
    


\subsection{MA-Fuzzing: Multi Attacker Fuzzing}
\begin{figure}[htpb]
	\centering
	\subfloat[Single attacker in SA-Fuzzing]{
		\fbox{\includegraphics[width=0.45\linewidth]{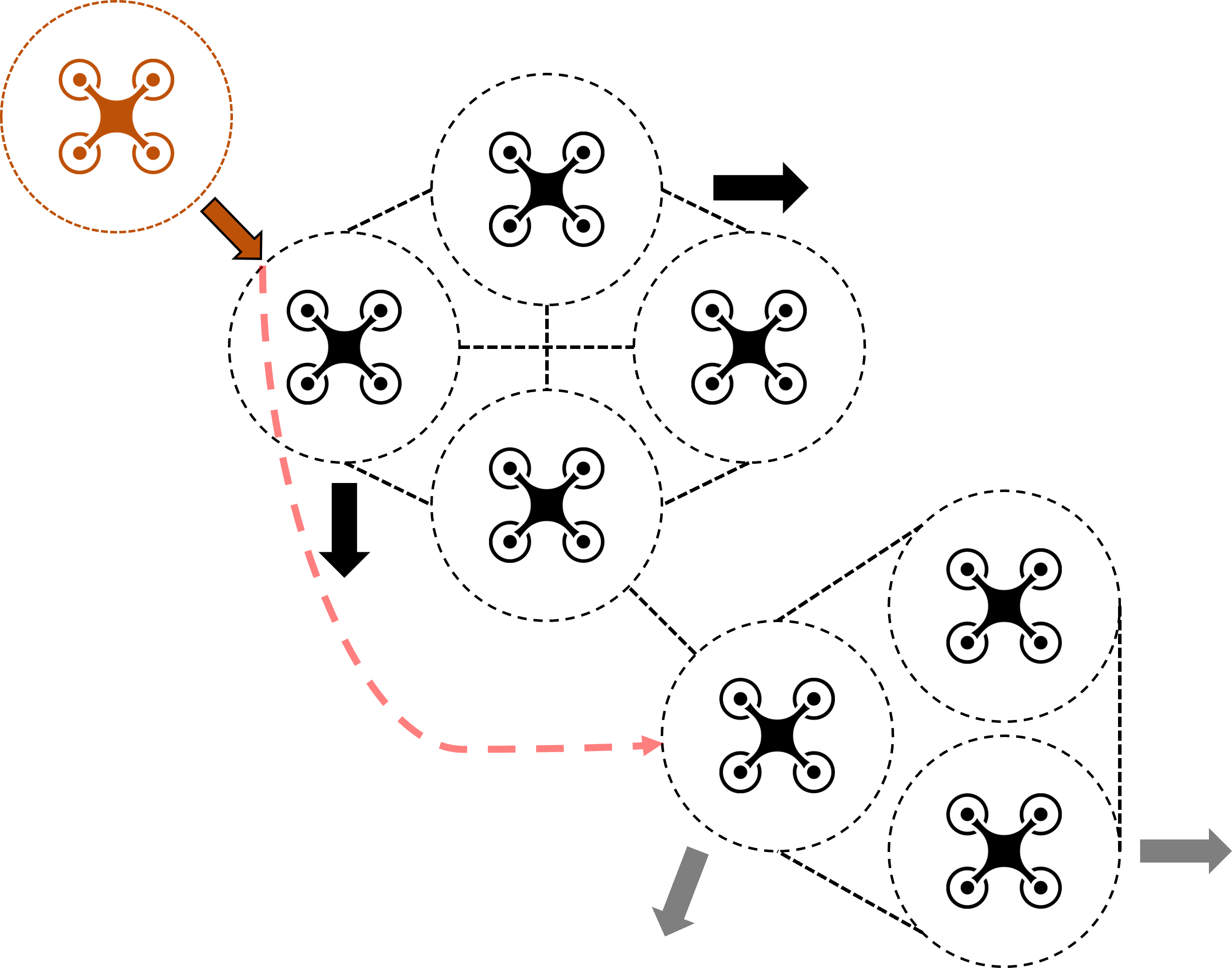}}
		\label{fig_6.1}
	}
	\subfloat[Multiple attackers in MA-Fuzzing]{
		\fbox{\includegraphics[width=0.45\linewidth]{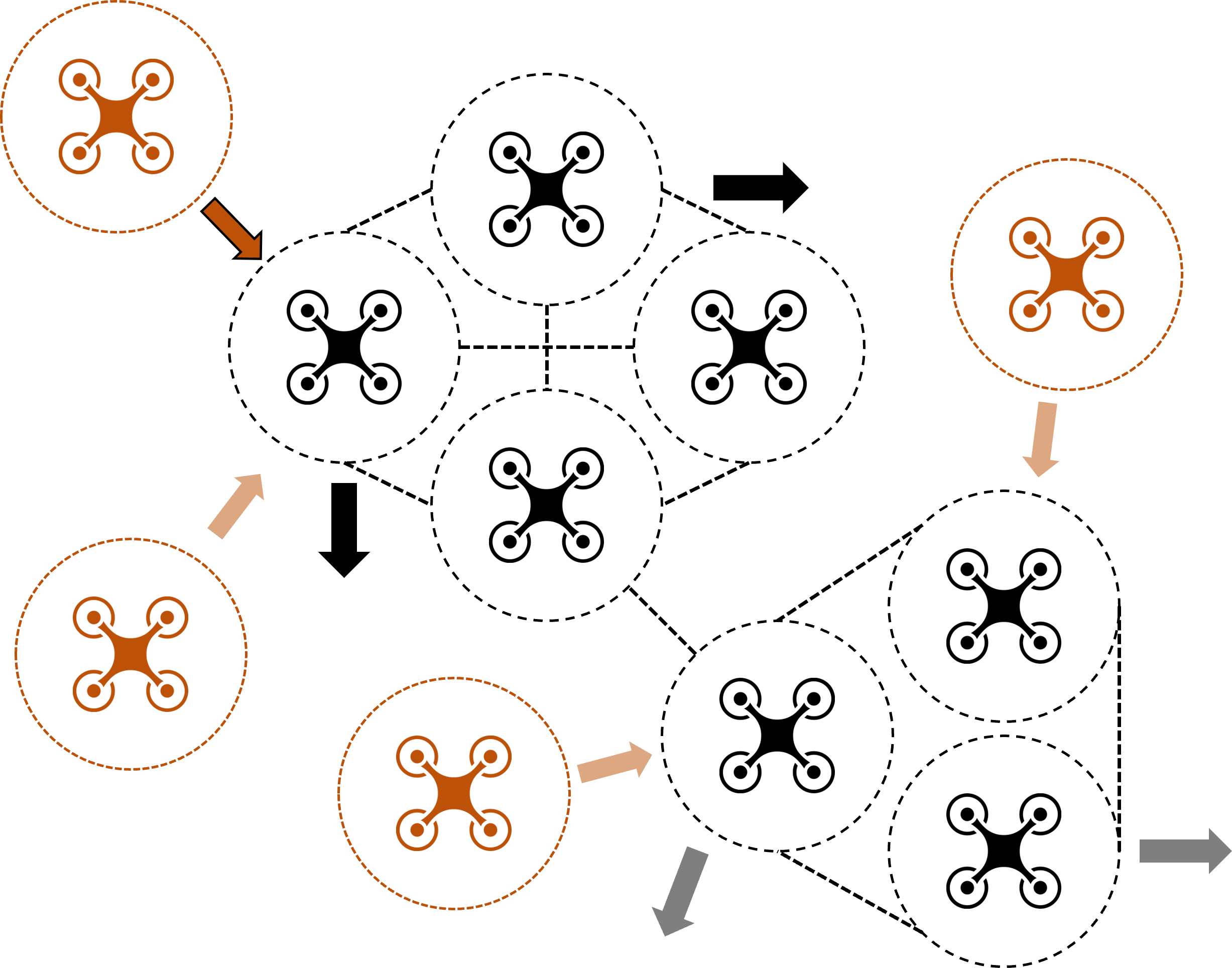}}
		\label{fig_6.2}
	}
	\caption{SA-Fuzzing refers to a testing process where there is only one attack drone. MA-Fuzzing involves multiple attack drones, but during the fuzzing process, only one attack drone is active at a time, while the others are considered nonexistent.}
	\label{fig_6}
\end{figure}

MA-Fuzzing is designed to comprehensively test the robustness of the swarm by leveraging multiple attack drones. At each iteration, one attack drone is activated while the others remain inactive, allowing focused perturbations. As illustrated in Fig. \ref{fig_6.2}, MA-Fuzzing identifies the key node from the entire swarm and samples candidate attack positions throughout the full sensing area of this global key node. This global perspective enables MA-Fuzzing to explore a wider range of failure-triggering scenarios by effectively guiding attack drone placement based on swarm-wide behavior.

MA-Fuzzing builds a constraint influence graph over the entire swarm. The key node $P_t$ is then selected globally using \textit{Katz} centrality.
Once the key node is identified, its sensing area is divided into $n$ equal-area sectors. MA-Fuzzing then randomly samples one candidate attack position from each sector, resulting in $n$ total candidate positions. MA-Fuzzing assumes that each attack drone can appear instantaneously at a new position, abstracting away motion continuity.
After generating all candidate positions, the robustness of the swarm is evaluated for each. If any drone is killed during execution, all test cases from the current execution are immediately output. Otherwise, the position yielding the minimum robustness is selected as the next attack position, and the fuzzing process proceeds to the next iteration.

\section{Evaluation}
\label{Evaluation}
In this section, we conduct multiple experimental evaluations to answer the following four research questions.
\begin{itemize}
	\item[$\bullet$] RQ1: Can RSFuzz detect logical vulnerabilities in swarm algorithms?
	\item[$\bullet$] RQ2: How effective is RSFuzz compared to SOTA?
	\item[$\bullet$] RQ3: How effective are the various components of RSFuzz?
    \item[$\bullet$] RQ4: What are the root causes of the logical vulnerabilities in swarm algorithms?
\end{itemize}
\subsection{Experiment Setup}
\begin{figure}[htpb]
    \setlength{\fboxsep}{0.5pt} 
	\centering
	\subfloat[Adaptive Swarm]{
		\fbox{\includegraphics[height=2.5cm,width=0.15\textwidth]{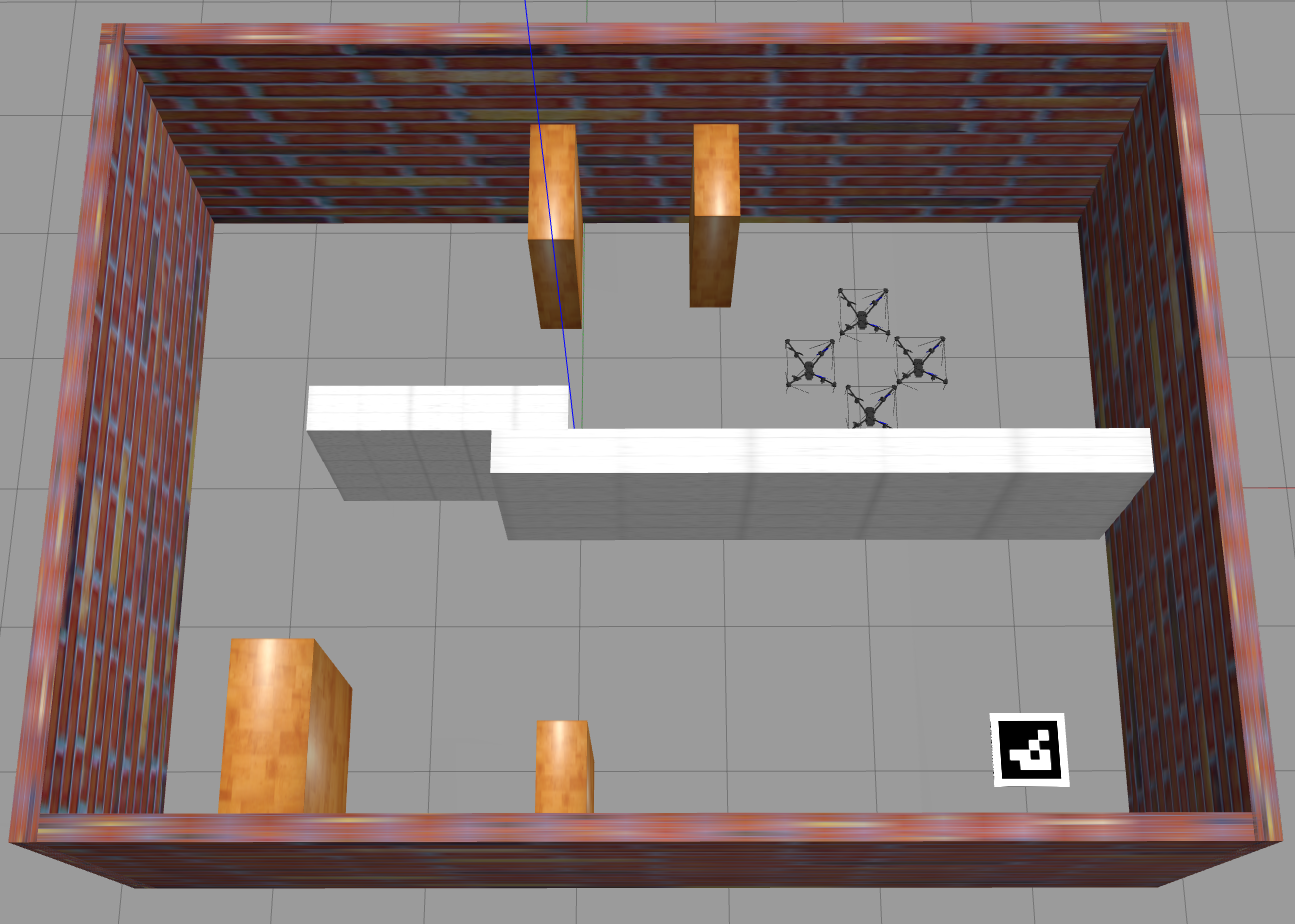}}
		\label{fig_4.1}
	}
	\subfloat[Pietro's]{
		\fbox{\includegraphics[height=2.5cm,width=0.15\textwidth]{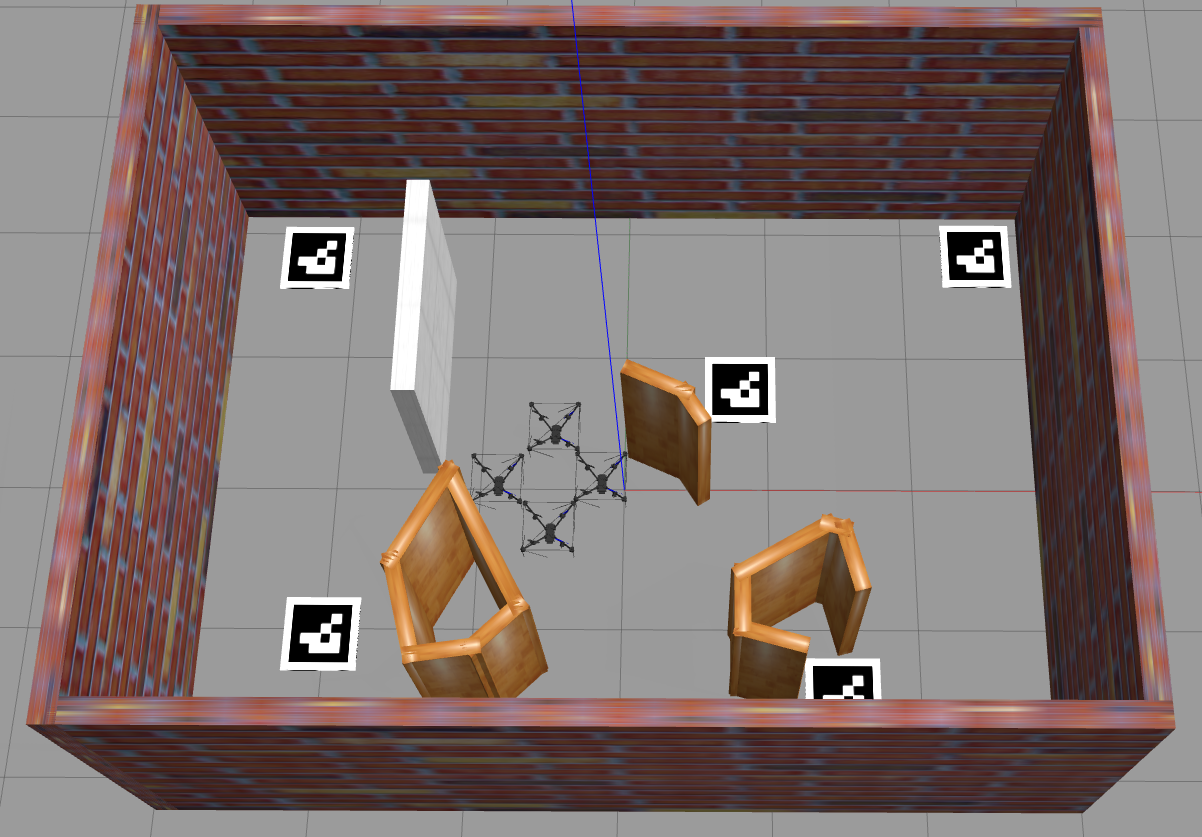}}
		\label{fig_4.2}
	}
	\subfloat[Howard's]{
		\fbox{\includegraphics[height=2.5cm,width=0.15\textwidth]{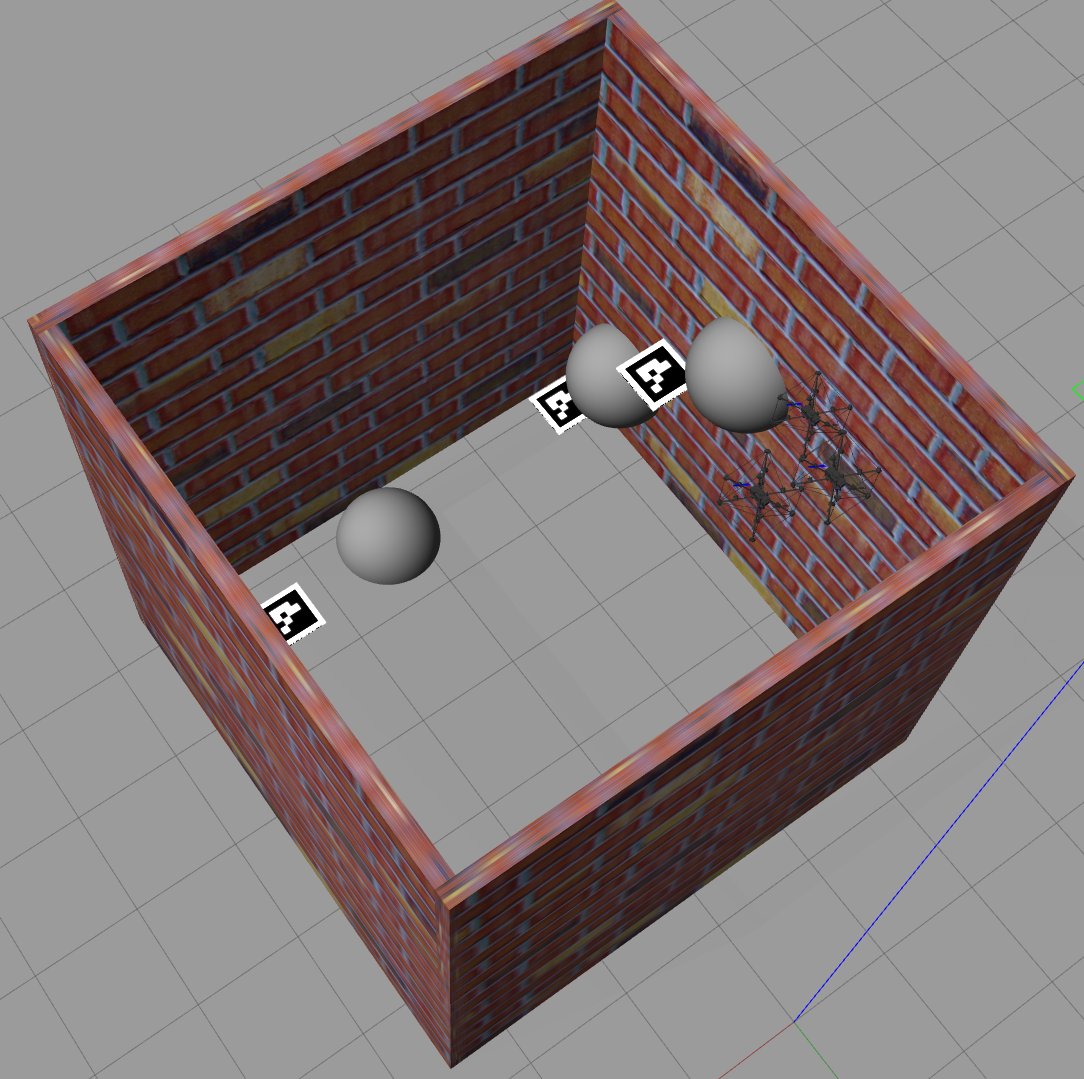}}
		\label{fig_4.3}
	}
	\caption{Visualizations of selected algorithms' missions.}
	\label{fig_4}
\end{figure}

\begin{table}[htpb]
        \caption{Selected Swarm Algorithms}
        \label{tab:table 1}
	\centering
 	\footnotesize
	\begin{tabular*}{\linewidth}{cccc}
		\toprule 
		\textbf{ID} & \textbf{Name} & \textbf{Language} & \textbf{Algorithm's Objective} \\
		\midrule 
		A1 & Adaptive Swarm \cite{agishev2019adaptive} & Python & Multi-agent navigation \\
		A2 & Pietro's \cite{carnelli2017SwarmRoboticsSim} & Matlab & Coordinated search and rescue \\
        A3 & Howard's \cite{howard2020swarm} & Matlab & Multi-agent navigation \\
		\bottomrule
	\end{tabular*}
\end{table}
\textbf {Target Swarm Algorithms. }To validate the effectiveness of the proposed scheme, we reviewed open-source research projects in swarm robotics from the past decade and selected those that were most suitable. The selection criteria for the program are as follows: \ding{172} It must be complete and executable; \ding{173} It should allow for the introduction of external objects, such as attack machines; \ding{174} The individuals in the swarm must exhibit cooperative behavior. To this end, we chose executable algorithms that exhibit collective swarm behaviors and allow us to introduce external objects. Table \ref{tab:table 1} presents the selected swarm algorithms. Fig. \ref{fig_4} shows visualizations of the swarm algorithms using the Gazebo simulator\cite{aguero2015inside}.

\begin{itemize}
\item[\textbf{A1}]Adaptive Swarm \cite{agishev2019adaptive} aims to move a swarm from the current position to a predefined destination while maintaining a formation and avoiding obstacles. The default swarm size is set to 4, with a maximum size of 20.
\item[\textbf{A2}]Pietro's algorithm\cite{carnelli2017SwarmRoboticsSim} aims to achieve cooperative rescue mission. The default swarm size is set to 10. The process is accelerated with more participating drones.
\item[\textbf{A3}]Howard's algorithm \cite{howard2020swarm} aims to move a swarm from the starting point to three destinations in a three-dimensional space, visiting the destinations in sequence, while maintaining formation and avoiding obstacles. The default swarm size is set to 4.
\end{itemize}

\begin{figure}[!htpb]
    \setlength{\fboxsep}{0.5pt} 
	\centering
	\subfloat[Drones colliding]{
		\fbox{\includegraphics[width=0.15\textwidth]{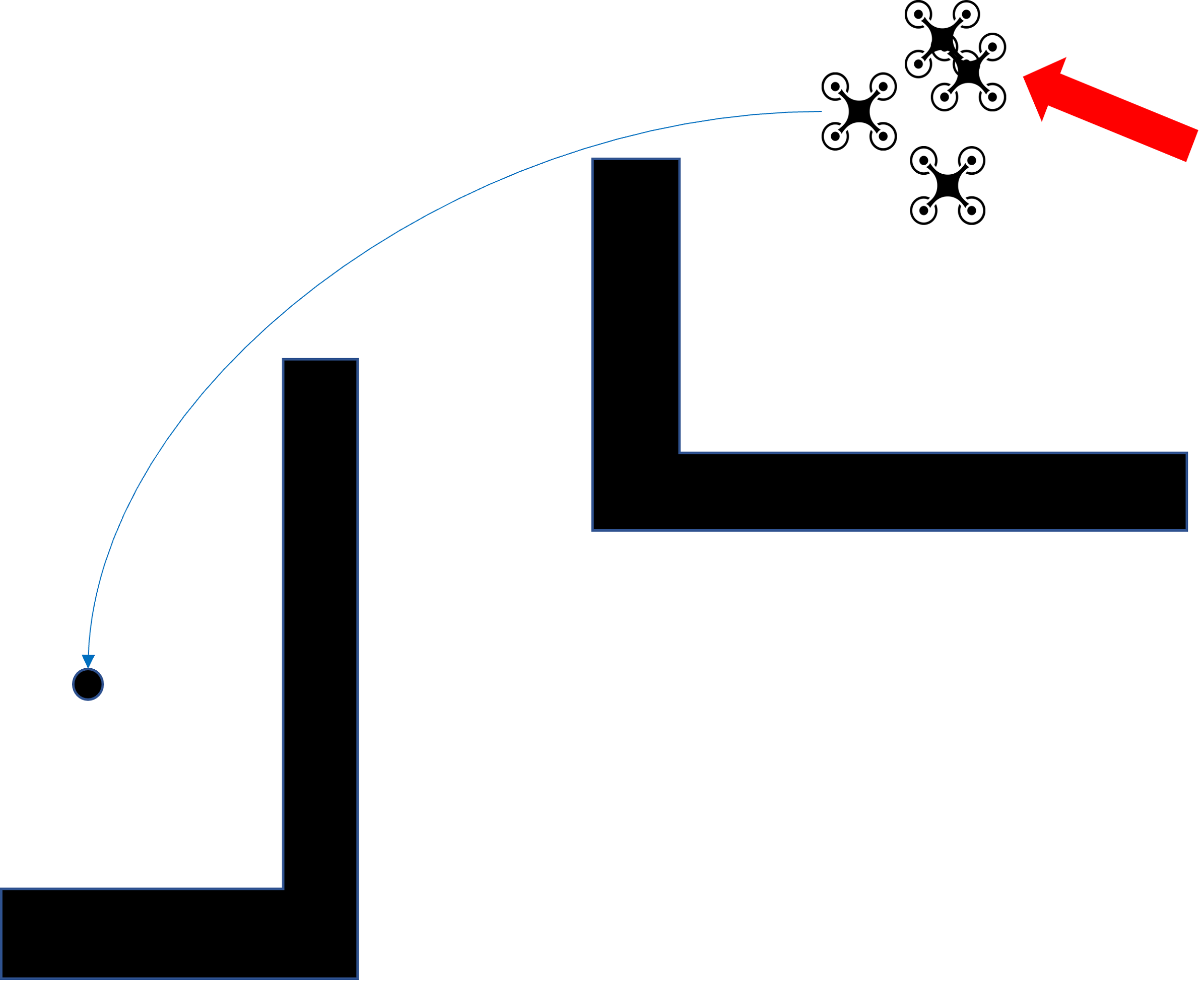}}
		\label{fig_3.1}
	}
	\subfloat[Obstacle crashes]{
		\fbox{\includegraphics[width=0.15\textwidth]{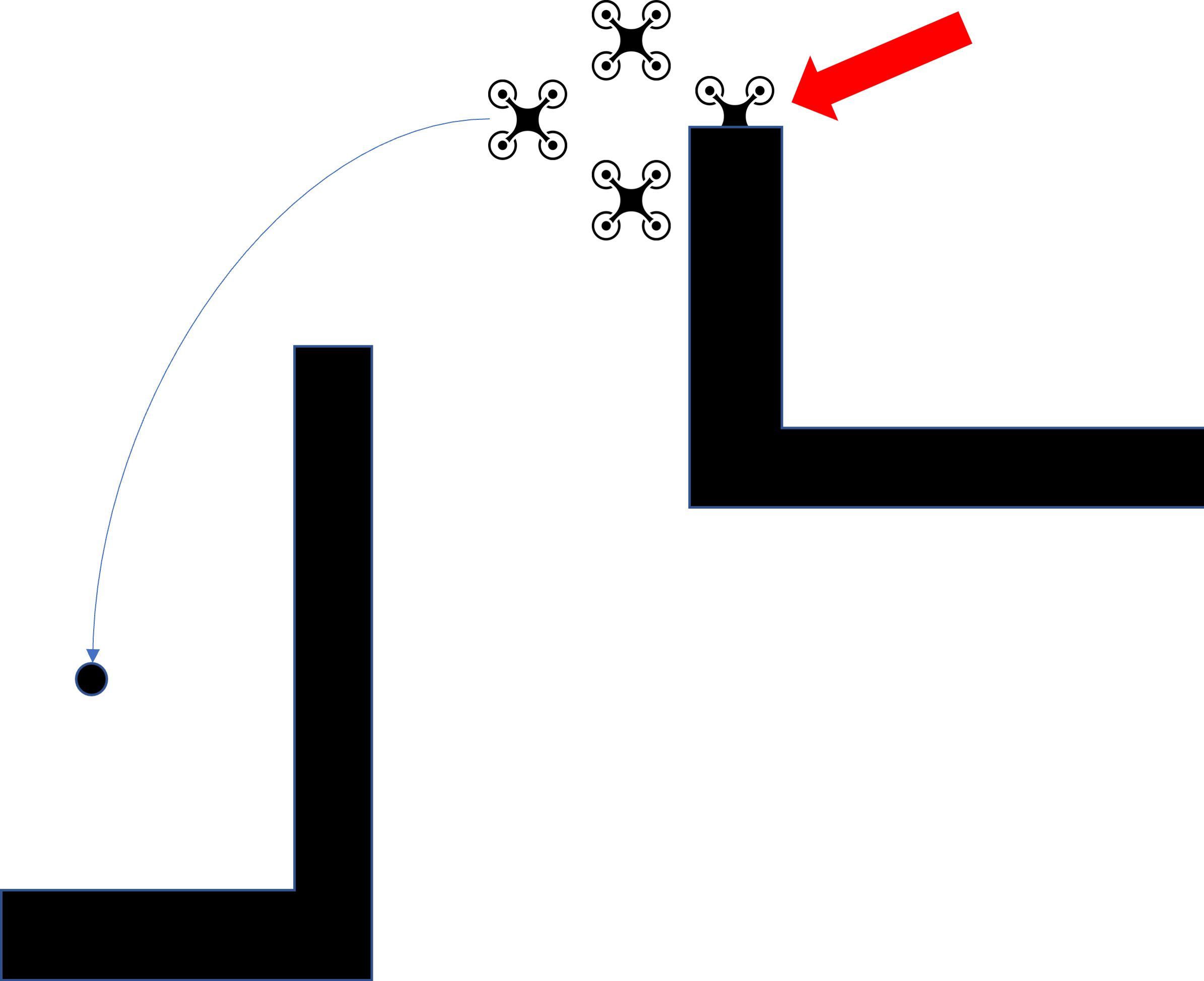}}
		\label{fig_3.2}
	}
	\subfloat[Timed out]{
		\fbox{\includegraphics[width=0.15\textwidth]{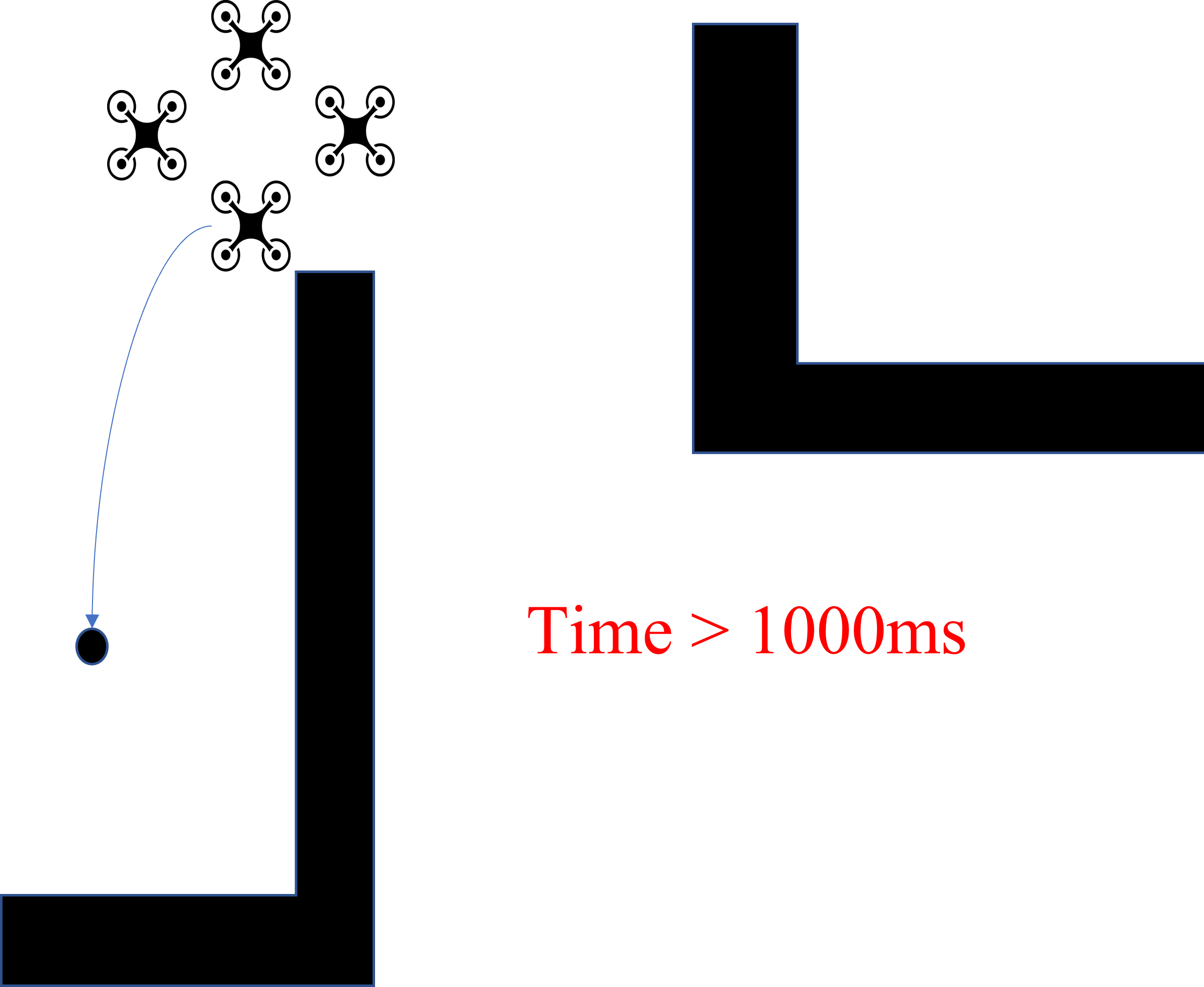}}
		\label{fig_3.3}
	}
	\caption{Mission failure scenarios.}
	\label{fig_3}
\end{figure}

\textbf {Experimental Setting. }We consider a swarm mission failed when: \ding{172} The swarm mission took more than twice as long to complete as its normal completion time; \ding{173} A drone in the swarm crashes into an obstacle; \ding{174} A drone in the swarm collides with another drone. Fig. \ref{fig_3} illustrates three sample scenarios in which a swarm mission fails. Fig. \ref{fig_3.1} shows the drones colliding with each other, Fig. \ref{fig_3.2} shows the swarm crashing into an obstacle, and Fig. \ref{fig_3.3} shows that more than 300 iterations were performed and timed out. It is important to note that we do not count attack drones crashing into the victim drone as a failure. Our attack drones are specifically designed to avoid direct collisions with victim drones.

\textbf {Implementation. }For our evaluation, we use a machine with Intel(R) i9-13900k 3.00Ghz and 128G RAM. The system executes Python-based programs on Ubuntu 22.04 and Matlab-based programs on Windows 10. 
\subsection{RSFuzz Evaluation Across Swarm Configurations (RQ1)}
\textbf{Programmes. }We evaluate RSFuzz on three algorithms (Algorithm A1, Algorithm A2, and Algorithm A3) as shown in Table \ref{tab:table 1}. To validate the effectiveness of RSFuzz at different mission sizes, we choose different numbers of swarm drones for algorithms A1, A2, and A3, respectively. For A1, we set the size of the swarms at 4, 6, and 8 drones, respectively. For A2, we set the size of the swarm to 10, 15, and 20 drones, respectively. For A3, we set the size of the swarms to 6, 8, and 10 drones, respectively. When constructing the constraint influence graph $G$, $D_{nominal}$ of A1 is \texttt{self.interrobots\_dist}, that of A2 is \texttt{r\_agent}, and that of A3 is \texttt{d\_min}.

\begin{figure*}[htpb]
	\centering
	\subfloat[RSFuzz for algorithm A1]{\includegraphics[width=0.25\textwidth]{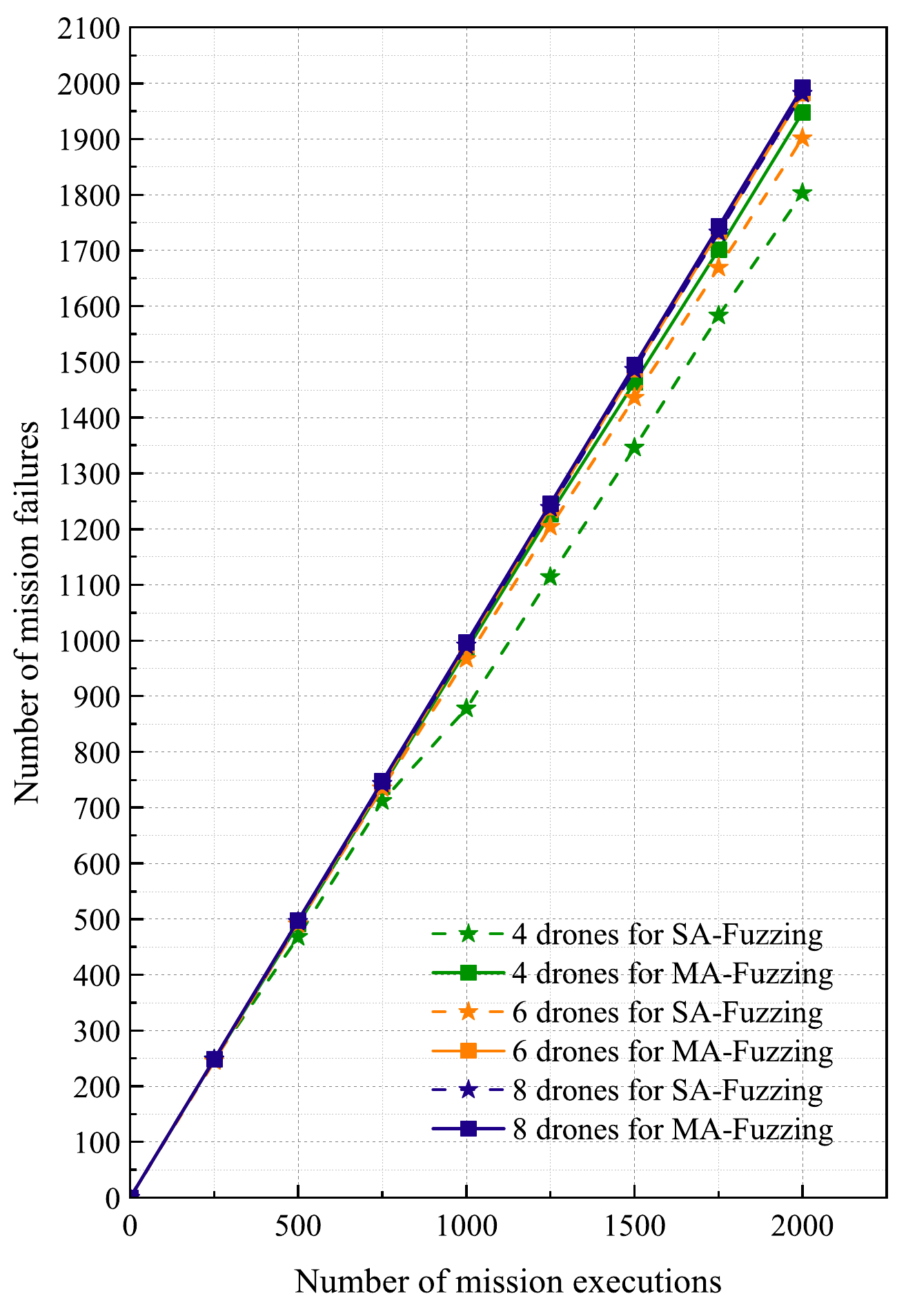}%
		\label{fig2.1}}
	\hfill
	\subfloat[RSFuzz for algorithm A2]{\includegraphics[width=0.25\textwidth]{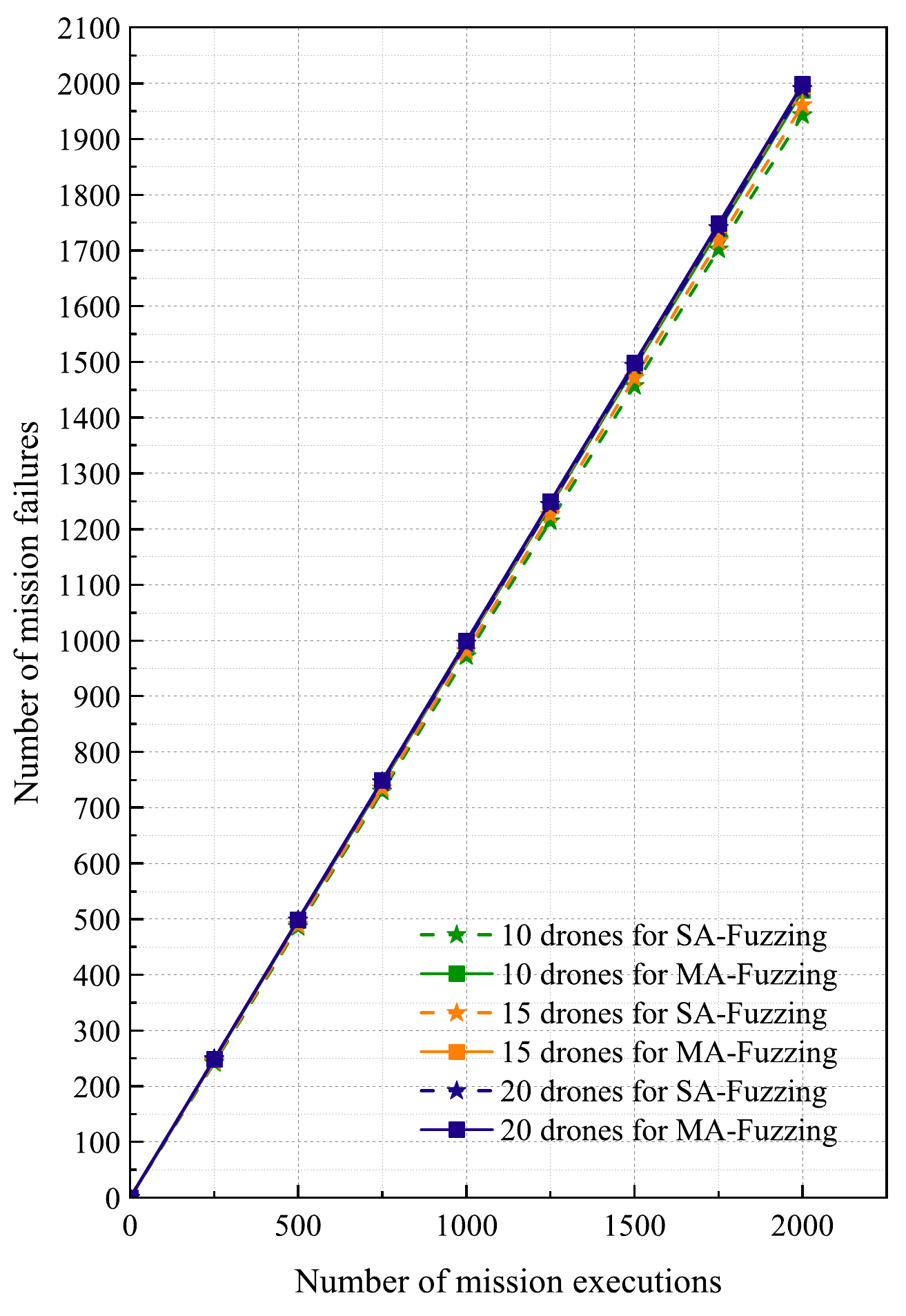}%
		\label{fig2.2}}
	\hfill
	\subfloat[RSFuzz for algorithm A3]{\includegraphics[width=0.25\textwidth]{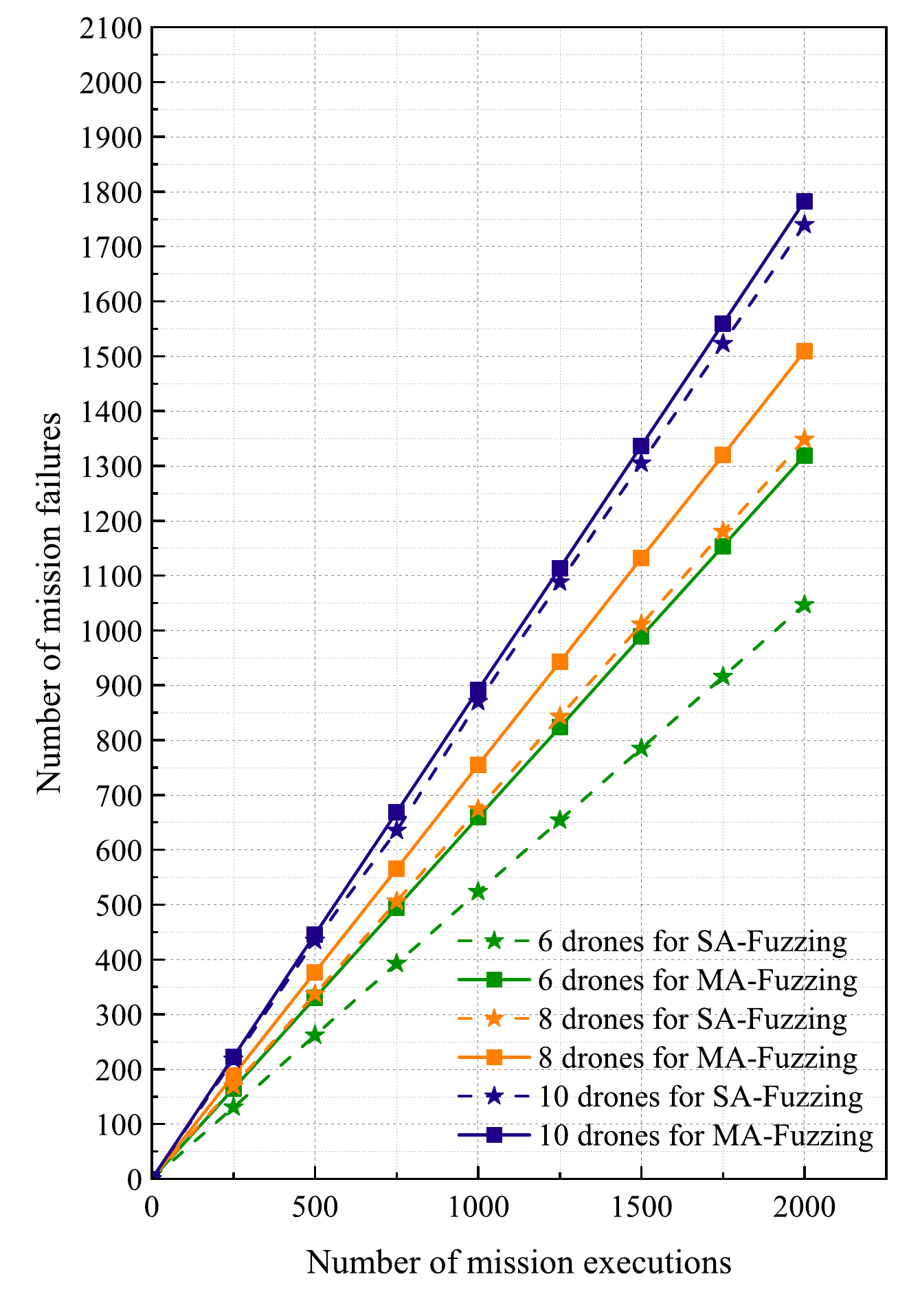}%
		\label{fig2.3}}

    \caption{Performance of RSFuzz in detecting vulnerabilities with different algorithms and swarm sizes}
	\label{fig2}
\end{figure*}

\textbf{Experimental Results. }The experimental results are shown in Table \ref{tab:table 2}, which presents the time cost of RSFuzz for vulnerability detection under different swarm sizes, as well as the failure rates of different algorithms. Column \textbf{Size} shows the various swarm sizes. Column \textbf{CTime} shows the average mission completion time, which we calculated by performing 50 swarm missions without introducing attack drones (i.e., without any interference). Column \textbf{Failure\textsubscript{w/o attack}} shows the mission failure rate for 50 missions without any perturbation. Column \textbf{Failure} shows the mission failure rate for 2000 missions. Algorithm A1 is a classical swarming algorithm, and the swarm failure rates are all over 90\%. The reasons for such high failure rates of algorithm A2 are the large swarm size, the complexity of the obstacles in the scene, and the high difficulty in completing the missions. For algorithm A3, the mission failure rate is not as high as the previous two because the default scenario of this algorithm is too simple. It can be seen that the success probability of detecting vulnerabilities in SA-Fuzzing and MA-Fuzzing is above 80\% on average. Fig. \ref{fig2} demonstrates the specific performance of RSFuzz in terms of its vulnerability detection. The success rate of detecting vulnerabilities in MA-Fuzzing is slightly higher than that of SA-Fuzzing. This higher success rate of MA-Fuzzing can be attributed to its ability to consistently identify the global robustness minimum.
\begin{table}[htpb]
    \caption{Results in Detecting Vulnerabilities}
    \label{tab:table 2}
	\centering
  	\footnotesize
	\begin{tabular*}{\linewidth}{@{\extracolsep{\fill}}cccccc}
		\toprule 
		\textbf{Method} & \textbf{Algo.} & \textbf{Size} & \textbf{Failure\textsubscript{w/o attack}} & \textbf{CTime(s)} & \textbf{Failure} \\
		\midrule 
		\multirow{10}{*}{\makecell{SA-\\Fuzzing}}&  & 4 & 2\% & 193.11 & 90.15\%\\
		& A1 & 6 & 2\% & 196.8 & 93.05\%\\
		&    & 8 & 4\% & 206.3 & 99.05\%\\\cmidrule(r){2-6}  
		&    & 10 & 4\% & 715.2 & 97.15\%\\
		& A2 & 15 & 4\% & 401.1 & 98.05\%\\
		&    & 20 & 2\% & 343.3 & 99.50\%\\\cmidrule(r){2-6}
		&    & 6 & 2\% & 17.28 & 52.35\%\\
		& A3 & 8 & 2\% & 22.50 & 67.4\%\\
		&    & 10 & 6\% & 27.27 & 87\%\\
		\midrule
		\multirow{10}{*}{\makecell{MA-\\Fuzzing}}&    & 4 & 2\% & 193.11 & 97.35\%\\
		& A1 & 6 & 2\% & 196.8 & 99.10\%\\
		&    & 8 & 4\% & 206.3 & 99.60\%\\\cmidrule(r){2-6}
		&    & 10 & 4\% & 715.2 & 99.35\%\\
		& A2 & 15 & 4\% & 401.1 & 99.70\% \\
		&    & 20 & 2\% & 343.3 & 99.90\%\\\cmidrule(r){2-6}
  		&    & 6  & 2\% & 17.28 & 65.95\%\\
		& A3 & 8 & 2\% & 22.50 & 75.45\%\\
		&    & 10 & 6\% & 27.27 & 89.15\%\\
		\bottomrule
	\end{tabular*}
\end{table}

An analysis of the results (Table~\ref{tab:table 2}) shows that, in most cases, the time required to detect vulnerabilities decreases as the swarm size increases. This trend may be attributed to the higher likelihood of collisions with obstacles in larger swarms. However, Algorithm A3 deviates from this pattern, likely because its scenarios are relatively simple and failures consistently occur at the same location. A detailed explanation of this behavior is provided in \nameref{Root Causes}. In general, RSFuzz shows strong effectiveness in detecting logical vulnerabilities in terms of both success rate and time efficiency.

It is important to highlight the practical significance of these failure events. Mission failures such as collisions, crashes, or timeout not only jeopardize the successful completion of critical tasks—ranging from battlefield reconnaissance to disaster relief—but may also result in costly hardware damage and potential safety hazards. The presence of attack drones exacerbates these risks by exploiting logical vulnerabilities in swarm coordination, triggering cascading failures. 

\subsection{Efficiency Comparison (RQ2)}
\textbf{Baseline. }To verify the performance of RSFuzz in detecting vulnerabilities in swarm algorithms, we used SWARMFLAWFINDER \cite{jung2022swarmflawfinder} as a baseline.

Algorithms A1 and A2 are evaluated in SWARMFLAWFINDER, and parts of algorithm A1 are open source. For Algorithm A2, we reimplemented SWARMFLAWFINDER based on the descriptions provided in the original paper, as the source code is not publicly available. 
We excluded SocraticSwarm \cite{henderson2018cost} (A2 in SWARMFLAWFINDER) and Sciadro \cite{cimino2019adaptive} (A3 in SWARMFLAWFINDER) due to critical reproducibility challenges. Specifically, SocraticSwarm's implementations depend on a legacy version of the Unity3D editor (2017). This version is no longer officially distributed via the Unity Hub, which only provides access to versions from 2020 and later. Sciadro also relies on the outdated NetLogo platform, making them infeasible to establish a stable and replicable execution environment for our experiments.

\textbf{Efficiency and Effectiveness Comparison. }A comparison between the two RSFuzz schemes, SA-Fuzzing and MA-Fuzzing, and SWARMFLAWFINDER is presented in Table \ref{tab:table 4}, focusing on both efficiency and effectiveness. Column \textbf{Algo.} indicates the algorithm used for testing. Column \textbf{Methods} lists the fuzzing methods evaluated, including SWARMFLAWFINDER (SFF), SA-Fuzzing, and MA-Fuzzing. Column \textbf{\#Failure} shows the number of mission failures detected out of 2000 mission executions. Column \textbf{Time Cost} represents the total time consumed during testing, formatted as hours:minutes:seconds. Column \textbf{Avg. time (s}) displays the average time in seconds to detect a vulnerability during successful runs. For algorithm A1, when the swarm size is 4, SA-Fuzzing improves the failure detection rate by 18.5\% and reduces average detection time by 10.3\%. MA-Fuzzing further boosts effectiveness by 27.9\% and lowers the average time by 60.9\%. 
\begin{table}[h!]
	\caption{Comparison of the Efficiency of RSFuzz and SWARMFLAWFINDER on A1 and A2 Algorithms}
        \label{tab:table 4}
	\centering
        \footnotesize
	\begin{tabular*}{\linewidth}{@{\extracolsep{\fill}}ccccc}
		\toprule 
		\textbf{Algo.} & \textbf{Methods} & \textbf{\#Failure} & \textbf{Time Cost} & \textbf{Avg. time(s)} \\
		\midrule 
		& SFF & 1521/2000 & 53:51:40 & 96.98 \\
		A1	& \textbf{SA-Fuzzing} & \textbf{1803/2000} & \textbf{43:34:12} & \textbf{86.99} \\
		& \textbf{MA-Fuzzing} & \textbf{1947/2000} & \textbf{20:31:48} & \textbf{37.95} \\
		\midrule
		& SFF & 1691/2000 & 82:20:54 &  151.175\\
		A2	& \textbf{SA-Fuzzing} & \textbf{1961/2000} & \textbf{67:27:36} & \textbf{123.85} \\
		& \textbf{MA-Fuzzing} & \textbf{1994/2000} & \textbf{29:47:24} & \textbf{53.781} \\
            \midrule
		&  SFF &  583/2000 &  7:52:15 &   14.168\\
		 A3	&  \textbf{SA-Fuzzing} &  \textbf{1047/2000} &  \textbf{7:21:33} &  \textbf{13.25} \\
		&  \textbf{MA-Fuzzing}&  \textbf{1319/2000} & \textbf{6:46:37} &  \textbf{12.199} \\
		\bottomrule
	\end{tabular*}
\end{table}
For algorithm A2, when the swarm size is 15, SA-Fuzzing and MA-Fuzzing achieve 15.97\% and 17.92\% higher detection rates, respectively, while reducing average detection time by 18.08\% and 64.4\%. For algorithm A3, with the swarm size set to 6, SA-Fuzzing improves the failure detection rate by a significant 79.6\% and shortens the average detection time by 6.5\%. MA-Fuzzing demonstrates even greater superiority, boosting the detection rate by an impressive 126.2\% while reducing the average time by 13.9\% compared to SFF.

\begin{table}[htpb]
  \centering 
  \caption{Comparison of logical vulnerabilities discovered by SWARMFLAWFINDER, SA-Fuzzing, and MA-fuzzing}
  \label{tab:overflow}
  \footnotesize 
  \begin{tabular}{@{} c l c c c @{}}
    \toprule
    \multirow{2}{*}{\textbf{ID}} & \multirow{2}{*}{\textbf{Mission Failure and Root Cause}} & \multicolumn{3}{c}{\textbf{Uniq. Vulns. Detected}} \\
    \cmidrule(l){3-5}
    & & \textbf{SFF} & \textbf{SA} & \textbf{MA} \\
    \midrule
    \multirow{18}{*}{{\textbf{A1}}} 
    & \cellcolor{tablegray}\textbf{Crash between victim drones} & \cellcolor{tablegray}\textbf{9} & \cellcolor{tablegray}\textbf{13} & \cellcolor{tablegray}\textbf{14} \\
    & \quad Missing collision detection & 8 & 8 & 8 \\
    & \quad Naive multi-force handling  & 4 & 4 & 4 \\
    & \quad Unsupported static movement & 1 & 1 & 1 \\
    & \quad Constraint conflicts & - & \textbf{3} & \textbf{4} \\
    & \quad Constraint overload & - & \textbf{1} & \textbf{1} \\
    \cmidrule(l){2-5}
 
    & \cellcolor{tablegray}\textbf{Crash into external objects} & \cellcolor{tablegray}\textbf{8} & \cellcolor{tablegray}\textbf{11} & \cellcolor{tablegray}\textbf{11} \\
    & \quad Missing collision detection & 3 & 3 & 3 \\
    & \quad Naive multi-force handling  & 3 & 3 & 3 \\
    & \quad Unsupported static movement & 1 & 1 & 1 \\
    & \quad Excessive force in APF      & 1 & 1 & 1 \\
    & \quad Constraint conflicts & - & \textbf{2} & \textbf{2} \\
    & \quad Constraint overload & - & \textbf{1} & \textbf{1} \\
    \cmidrule(l){2-5}
 
    & \cellcolor{tablegray}\textbf{Suspended progress} & \cellcolor{tablegray}\textbf{2} & \cellcolor{tablegray}\textbf{-} & \cellcolor{tablegray}\textbf{-} \\
    & \quad Naive swarm's pose measurement & 1 & - & - \\
    & \quad Insensitive object detection   & 1 & - & - \\
    \cmidrule(l){2-5}
 
    & \cellcolor{tablegray}\textbf{Slow progress} & \cellcolor{tablegray}\textbf{1} & \cellcolor{tablegray}\textbf{-} & \cellcolor{tablegray}\textbf{-} \\
    & \quad Insensitive object detection & 1 & - & - \\
    \midrule
    
    \multicolumn{2}{l}{\textbf{Total Unique Vulnerabilities}} & \textbf{20} & \textbf{24} & \textbf{25} \\
 \midrule
 \multirow{7}{*}{{\textbf{A2}}} 
 
    & \cellcolor{tablegray}\textbf{Crash into external objects} & \cellcolor{tablegray}\textbf{3} & \cellcolor{tablegray}\textbf{4} & \cellcolor{tablegray}\textbf{5} \\
    & \quad Naive Detouring method & 1 & 1 & 1 \\
    & \quad Detouring without sensing  & 2 & 2 & 2 \\
    & \quad Vuln. of uniform coverage mechanism   & - & \textbf{1} &  \textbf{2} \\
    \cmidrule(l){2-5}
 
    & \cellcolor{tablegray}\textbf{Slow progress} & \cellcolor{tablegray}\textbf{2} & \cellcolor{tablegray}\textbf{3} & \cellcolor{tablegray}\textbf{4} \\
    & \quad Insensitive object detection & 2 & 2 & 2 \\
    & \quad Vuln. of uniform coverage mechanism   & - & \textbf{1} & \textbf{2} \\
    \midrule
    
    \multicolumn{2}{l}{\textbf{Total Unique Vulnerabilities}} & \textbf{5} & \textbf{7} & \textbf{9} \\
    \midrule
 \multirow{7}{*}{{\textbf{A3}}}
 & \cellcolor{tablegray}\textbf{Crash between victim drones} & \cellcolor{tablegray}\textbf{1} & \cellcolor{tablegray}\textbf{2} & \cellcolor{tablegray}\textbf{3} \\
    & \quad The gradient descent obstacle avoidance & 1  & \textbf{2} & \textbf{3}\\
    & \quad mechanism is too simple & & \\ 
    \cmidrule(l){2-5}
  & \cellcolor{tablegray}\textbf{Crash into external objects} & \cellcolor{tablegray}\textbf{1} & \cellcolor{tablegray}\textbf{2} & \cellcolor{tablegray}\textbf{4} \\
    & \quad The gradient descent obstacle avoidance & 1  & \textbf{2} & \textbf{4}\\
    & \quad mechanism is too simple & & \\ 
    \midrule
    \multicolumn{2}{l}{\textbf{Total Unique Vulnerabilities}} & \textbf{2} & \textbf{4} & \textbf{7} \\
    \bottomrule
  \end{tabular}
\end{table}
To evaluate bug detection capabilities, we performed a direct comparison between RSFuzz (and its variants) and SWARMFLAWFINDER (SFF) across algorithms A1, A2, and A3. As shown in Table \ref{tab:overflow}, which uses root causes found by SFF as a baseline, RSFuzz surpasses SFF's detection scope.

We observed that in tests on algorithm A1, our methods did not detect timeout vulnerabilities (e.g., "Suspended progress"). This is a direct consequence of the dense obstacle environment in A1. Because our robustness score is computed from several rules, the prominent "distance to obstacles" rule guided the fuzzer to prioritize discovering crash-related vulnerabilities. As a result, drones were induced to crash before timeout conditions could be met.

This configurability in RSFuzz's guidance mechanism offers the potential to tailor the fuzzing process. For example, by increasing the weight of the "progress towards target" rule within the robustness score, the fuzzer could be steered to investigate specific failure modes, such as mission timeouts.

Overall, both schemes of RSFuzz outperform SWARMFLAWFINDER in terms of effectiveness and efficiency, with MA-Fuzzing achieving the best results.
\subsection{Ablation Study (RQ3)}
We configured the swarm size to four drones and conducted an ablation study using algorithm A1 to evaluate the individual and combined contributions of RSFuzz's two key components: (A) key node identification using \textit{Katz} centrality, and (B) robustness-guided attack drone positioning. Four fuzzing variants were evaluated: \ding{172} A-B- (Random Fuzzing), where both components are disabled—a target drone is randomly selected, and attack drones are randomly placed around it; \ding{173} A+B-, which enables key node identification but places attack drones randomly; \ding{174} A-B+, which disables key node identification but uses robustness metrics to guide attack drone placement; and \ding{175} A+B+ (SA-Fuzzing). Each variant was evaluated under three different drone perception radii: 0.10m, 0.15m, and 0.20m.

\begin{figure}[htpb]
    \centering
    \includegraphics[width=0.8\linewidth]{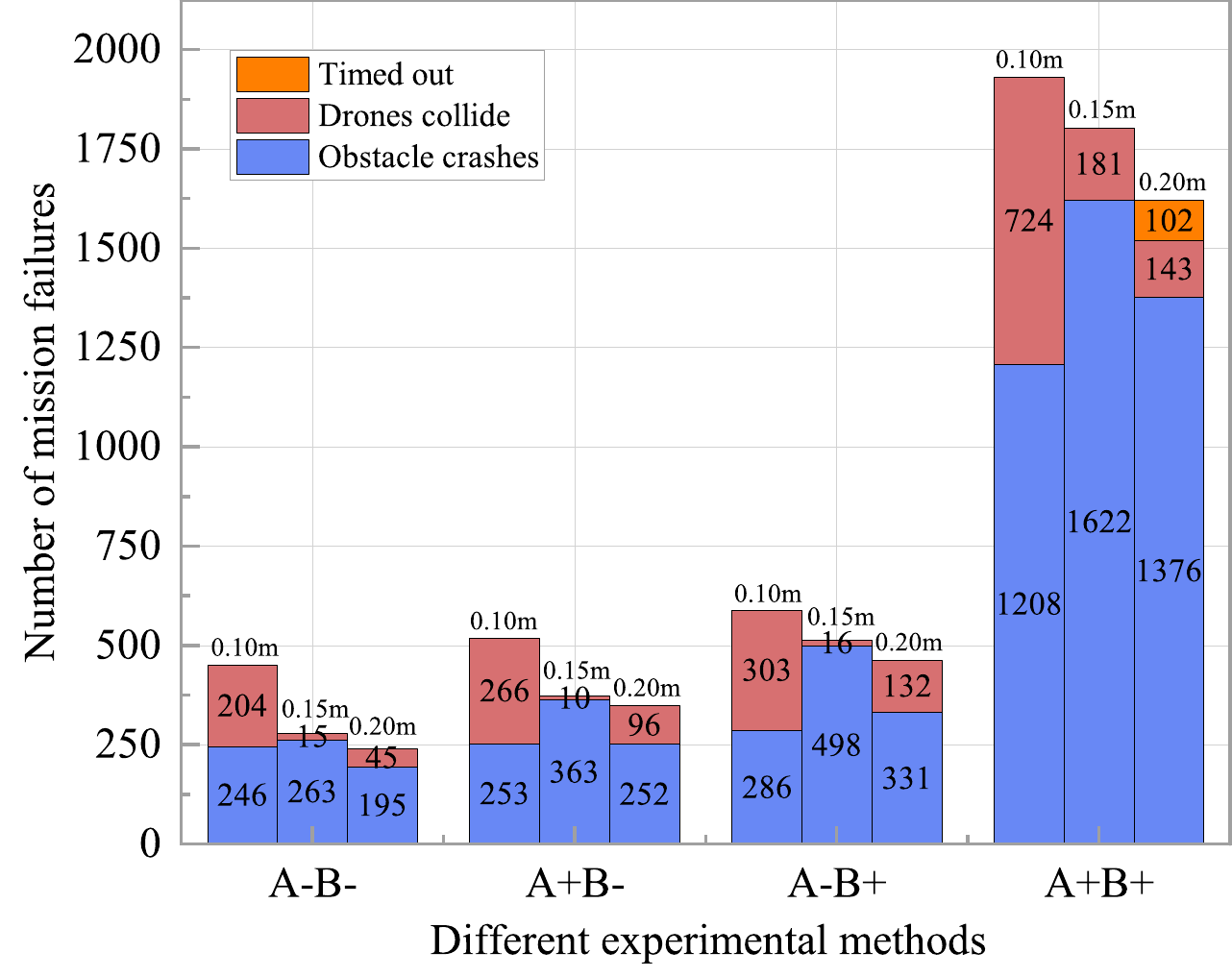}
    \caption{Results of ablation experiments}
    \label{fig3}
\end{figure}

The default value of the radius of the potential field of the drone in Adaptive Swarm is 0.15m, i.e., the area repelled by the drone near an obstacle is a circular potential field with a radius of 0.15m. As shown in Fig. \ref{fig3}, in each test method, the leftmost, central, and rightmost bars within the group correspond to drone potential field radii of 0.10m, 0.15m, and 0.20m, respectively, with all other parameters held constant across trials. At the default potential field radius of 0.15m (Group B), SA-Fuzzing triggers a significantly higher number of mission failures—1803 out of 2000 executions—compared to 278, 373, and 514 failures for random fuzzing, key-drone-only fuzzing, and robustness-guided fuzzing, respectively.
\begin{itemize}
\item[$\bullet$]Introducing a key drone increases the number of mission failures by approximately 34.2\% compared to random fuzzing (from 278 to 373 failures).
\item[$\bullet$]Applying robustness-guided attacker positioning increases mission failures by approximately 84.9\% compared to random fuzzing (from 278 to 514 failures).
\item[$\bullet$]Combining both strategies in SA-Fuzzing leads to a total increase of approximately 548.9\% compared to random fuzzing (from 278 to 1803 failures).
\end{itemize}

And, when the radius is reduced to 0.10 m, the success rate of fuzzing increases for all three methods; however, the percentage of collisions between drones leading to mission failure is significantly increased at 0.10 m potential field radius compared to 0.15 m. This is because when the radius of the potential field is small, the swarm may tend to go to a narrower route, resulting in a dense swarm and an increased probability of collision between drones during the mission.

Finally, the fuzzing efficiency of all three methods decreases when the radius of the potential field is increased to 0.20 m. This is because a larger radius of the potential field reduces the probability of the drone colliding with an obstacle. However, the larger potential field radius causes the swarm to make obstacle avoidance too early, which leads to the SA-Fuzzing method triggering a new type of mission failure: Timed out.

Overall, the results demonstrate that introducing key drone and robustness guidance strategies significantly enhances the ability of fuzzing techniques to detect failures, with SA-Fuzzing consistently achieving the highest mission failure rates across different potential field radii.

\begin{table}[htpb]
    \caption{Reasons for Failure of Swarm Missions}
        \label{tab:table 3}
	\centering
	\footnotesize
	\begin{tabular*}{\linewidth}{@{\extracolsep{\fill}}ccccc}
		\toprule 
		\textbf{Alg.} & \textbf{Size} & \textbf{Mission Failure Reasons} & \textbf{Occurrences} & \textbf{Percentage} \\
		\midrule
		&   & Drones colliding & 181/1803 & 10.03\%\\ 
		& 4 & Obstacle crashes & 1622/1803 & 89.94\%\\
		&	& Timed out & - & -\\
		\cmidrule(r){2-5}
		&   & Drones colliding & 645/1901 & 33.93\%\\ 
		A1  & 6	& Obstacle crashes & 1256/1901 & 66.07\%\\
		&	& Timed out & - & -\\
		\cmidrule(r){2-5} 
		&   & Drones colliding & 1339/1981 & 67.59\%\\
		& 8	& Obstacle crashes & 642/1981 & 32.41\%\\
		&	& Timed out & - & -\\
		\midrule
		& \multirow{2}{*}{10} & Obstacle crashes & 1583/1943 & 81.48\%\\
		&	 & Timed out & 360/1943 & 18.52\%\\
		\cmidrule(r){2-5}
		A2  &  \multirow{2}{*}{16}  & Obstacle crashes & 1794/1961 & 91.48\%\\
		&	 & Timed out & 167/1961 & 8.52\%\\
		\cmidrule(r){2-5} 
		&  \multirow{2}{*}{20}  & Obstacle crashes & 1863/1990 & 93.59\%\\
		&	 & Timed out & 127/1990 & 6.41\%\\
  		\midrule
		&   & Drones colliding & 800/1047 & 76.41\%\\ 
		& 6 & Obstacle crashes & 247/1047 & 23.59\%\\
		&	& Timed out & - & -\\
		\cmidrule(r){2-5}
		&   & Drones colliding & 760/1348 & 56.38\%\\ 
		A3  & 8	& Obstacle crashes & 588/1348 & 43.62\%\\
		&	& Timed out & - & -\\
		\cmidrule(r){2-5} 
		&   & Drones colliding & 1177/1740 & 67.64\%\\
		& 10& Obstacle crashes & 563/1740 & 32.36\%\\
		&	& Timed out & - & -\\
		\bottomrule
	\end{tabular*}
\end{table}
\subsection{Root Causes (RQ4)}
\label{Root Causes}
The root causes of swarm mission failures across different algorithms (A1–A3) and swarm sizes are summarized in Table \ref{tab:table 3}. The column \textbf{Alg.} indicates the algorithm used, while \textbf{Size} represents the number of drones in the swarm. The column \textbf{Mission Failure Reasons} lists the types of failures observed, including drone collisions, obstacle crashes, and mission timeouts. \textbf{Occurrences} is the number of times each failure occurred, and \textbf{Percentage} shows the corresponding proportion relative to the total number of mission failures under that configuration.

\textbf{Fundamental vulnerability of artificial potential field in multi-constraint environment in Algorithm A1.} Table \ref{tab:table 3} reveals algorithm A1 collision rate surges from 10.03\% to 67.39\% as the swarm size increases (4→8), precisely because its artificial potential filed mechanism optimizes for single safety constraint at a time. This inherent design vulnerability manifests in two fatal patterns:
\begin{itemize}
    \item Constraint Conflicts: When the swarm (4 drones) needs to avoid both the attack drone and the obstacle, the individual constraints will conflict, and the local optima of potential field gradients force drones into collisions
    \item Constraint Overload: In the extreme case of 6-8 drones, the inter-agent safety constraints (internal constraints) exceed the algorithm's computational capacity, and the attack drone may trigger a chain collision.
\end{itemize}

\textbf{Incomplete safety constraint architecture in Algorithm A2.} For algorithm A2, an unusual phenomenon was observed: in the first 200 experiments, all mission failures were due to inter-drone collisions. Analysis of the A2 source code revealed that all drones initially depart from a single point (i.e., the base), resulting in inevitable collisions before full dispersion. Even after complete dispersion, over 95\% of failures remained due to inter-drone collisions. Further analysis showed that the original algorithm does not treat inter-drone collisions as fatal and thus lacks an avoidance mechanism for such events. As shown in Table \ref{tab:table 3}, the final experimental results of algorithm A2 indicate that, regardless of the swarm size, the primary cause of mission failure is drone collision with obstacles. Furthermore, the probability of drones colliding with obstacles increases as the swarm size grows (10→15→20). By analyzing the source code, this algorithm has two main vulnerabilities:
\begin{itemize}
    \item Detouring without sensing: When avoiding attack drones, the victim drone does not check for obstacles in its avoidance path.
    \item Vulnerability of uniform coverage mechanism: When the swarm spreads out to search, if there are already drones in a certain direction, the remaining drones are instructed to move in other directions. If the attacker affects the flight direction of one of the drones, it will trigger a chain reaction, causing other drones to urgently change their flight directions, which may conflict with the obstacle avoidance constraints and cause collisions with obstacles or timeout failure\footnote{For algorithm A2, the original design does not consider drone-to-drone collisions as a failure condition and omits inter-drone avoidance mechanisms. Accordingly, in our experiments, constraints related to inter-drone distance are not included in the robustness calculation, and $rob_4$ is excluded from $\mathbb{R}$.} (While Algorithm A2 implements directional avoidance, this behavior merely reflects a coverage-driven task allocation strategy, not an active inter-drone collision prevention mechanism). 
\end{itemize}

\begin{figure}[htpb]
	\centering
	\includegraphics[width=\linewidth]{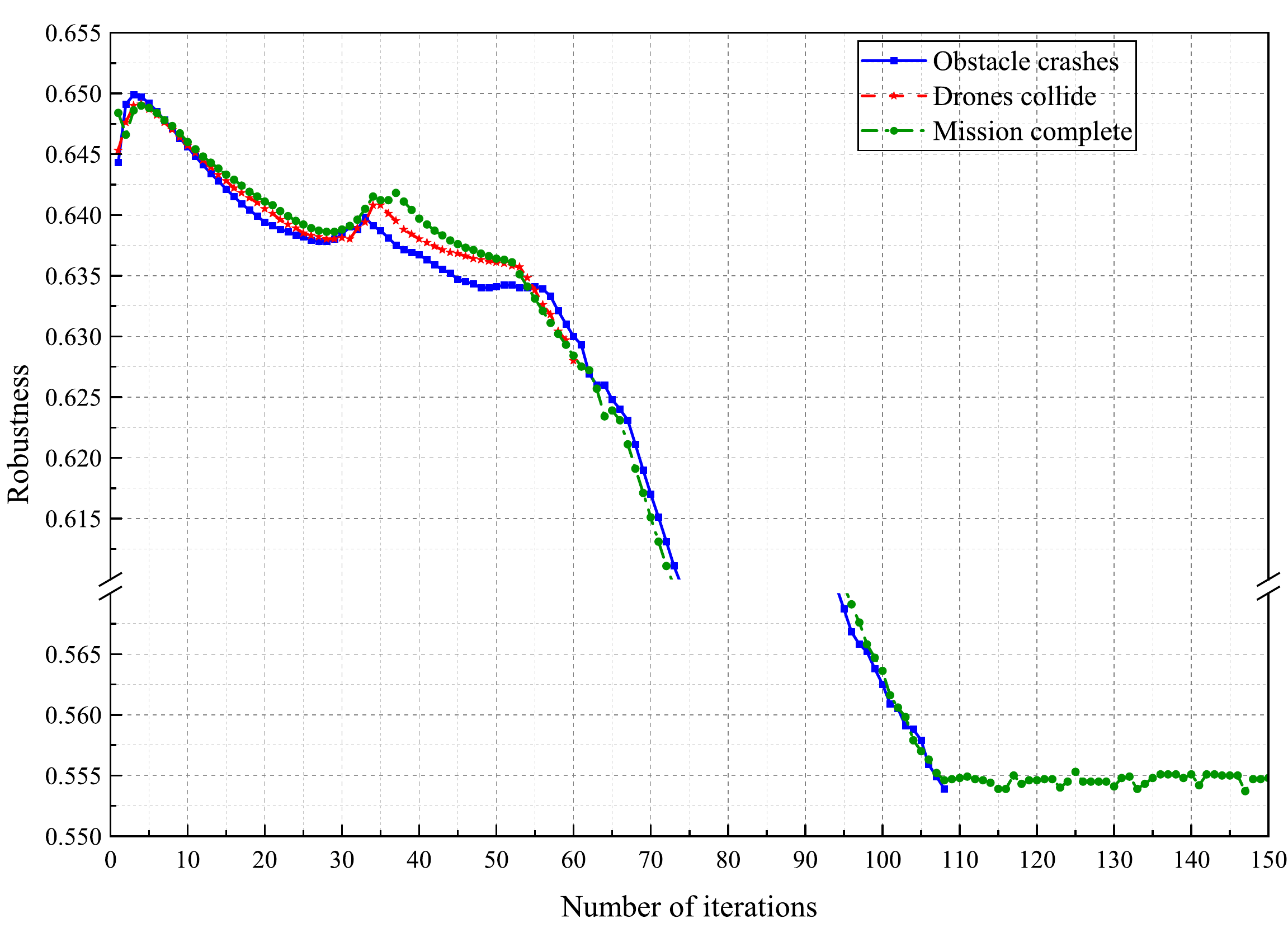}
	\caption{Swarm robustness variation with number of iterations}
	\label{Robustness}
\end{figure}

\textbf{Oversimplified Obstacle Avoidance in Algorithm A3.} Experimental results for algorithm A3 indicate that the time required to detect vulnerabilities doesn't decrease as the swarm size increases. Code analysis and robustness degradation patterns (Fig. \ref{Robustness}) reveal a critical limitation in Algorithm A3:
\begin{itemize}
    \item The Gradient Departure Obstacle Avoidance Mechanism is too Simple: The target exerts constant attraction to guide swarm movement, while obstacles generate proximity-dependent repulsion. During the initial phase (iterations 10-60), weaker target attraction allows obstacle repulsion to dominate, enabling safe navigation. However, upon reaching the second target point and executing the mandatory right-angle turn, surging target attraction overpowers repulsive forces. This imbalance compromises obstacle avoidance capabilities as drones prioritize the shortest path to the target, ultimately causing collisions with obstacles or other drones. 
\end{itemize}

Consequently, in the experiments with Algorithm A3, the swarm prioritizes reaching the target, and no mission failures were caused by timeouts. The experimental results in Table \ref{Robustness} show that the average time for Algorithm A3 to detect vulnerabilities is approximately half of the average mission completion time. From this analysis, it can be concluded that mission failures primarily occur between iterations 60 and 105, which is around the midpoint of the mission (iteration 150).
\subsection{Case Study}
\begin{figure}[htpb]
	\centering
	\includegraphics[width=\linewidth]{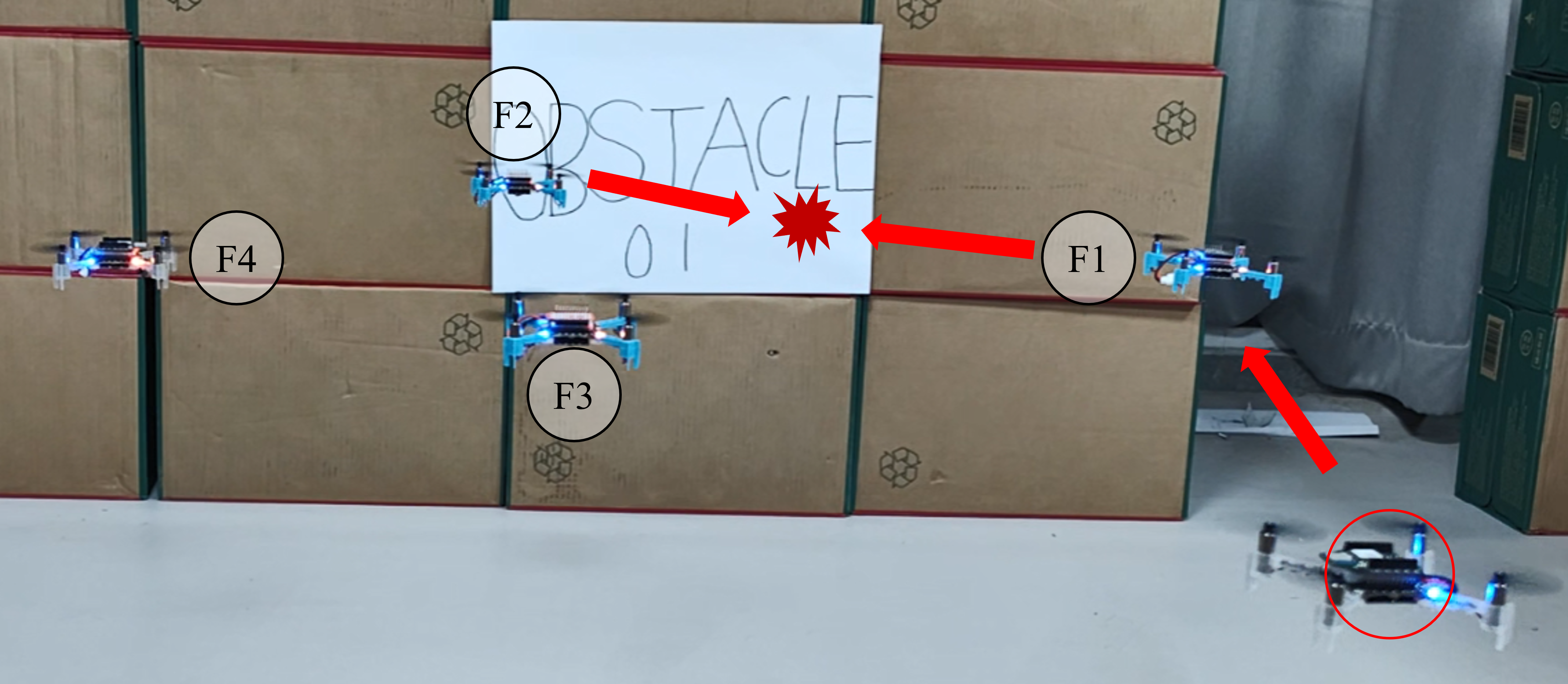}
	\caption{The attack drone (the red circle) approaches the key node drone (F1), causing it to move closer to another drone (F2) positioned near an obstacle. This forces F2 into a conflicting situation where it must simultaneously satisfy the behavioral constraints of obstacle avoidance and safe inter-drone spacing. As a result, F1 and F2 ultimately collide.}
	\label{attack_final}
\end{figure}

\begin{figure}[htpb]
	\centering
	\includegraphics[width=\linewidth]{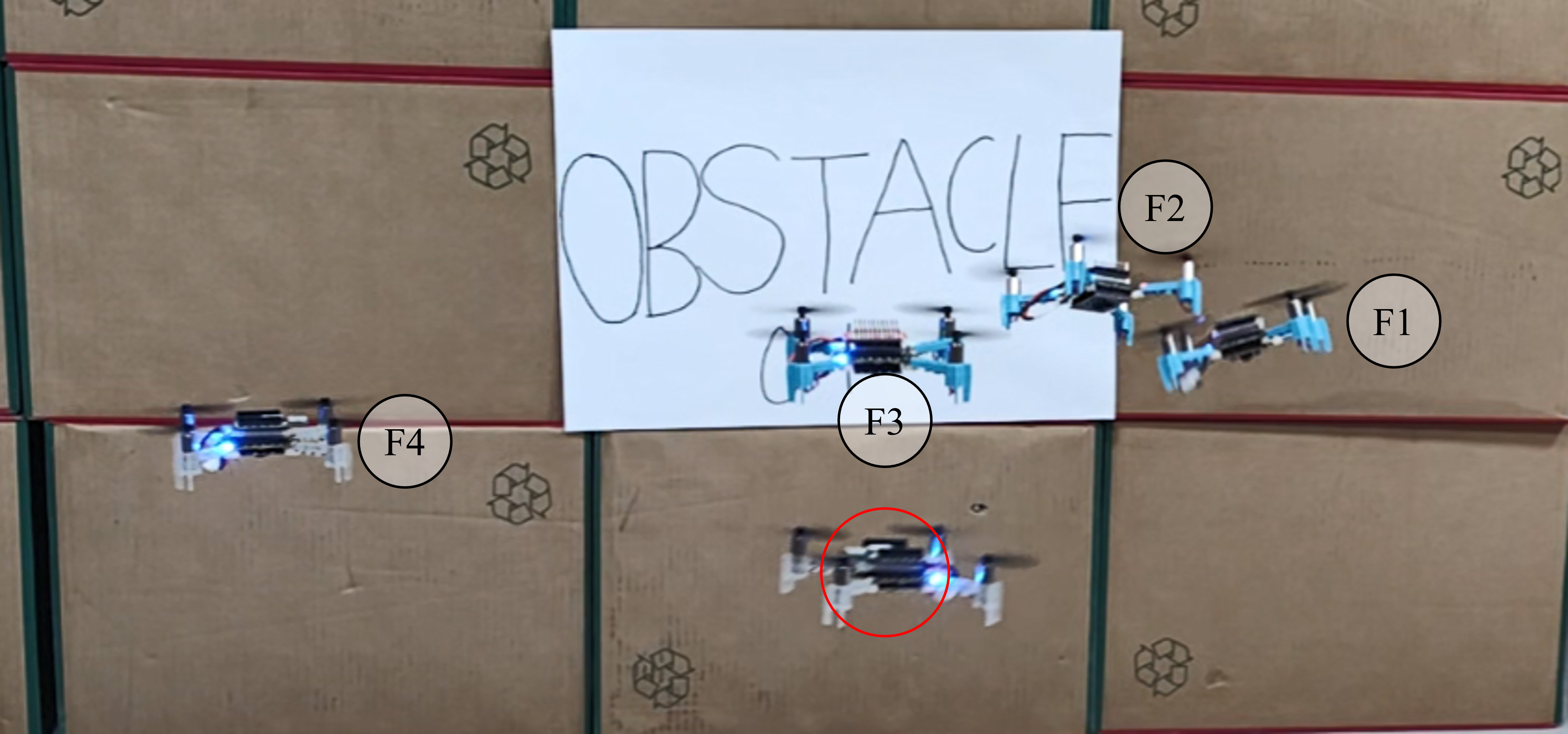}
	\caption{The collision between drones F1 and F2 captured during real-world.}
	\label{crush}
\end{figure}

We validated some of the vulnerabilities discovered in Algorithm A1 in real-world scenarios (Fig. \ref{attack_final} and Fig. \ref{crush}).

Constraint Conflicts: Fig. \ref{attack_final} shows a critical vulnerability in Algorithm A1 when faced with multiple safety constraints. The F2 drone was unable to simultaneously handle the constraints of avoiding the attacking drone and maintaining formation, ultimately leading to mission failure.

\textbf{Analysis.} This vulnerability originates from the architectural disconnect between two key components in Algorithm A1: the formation planner \texttt{formation} and the path planner \texttt{local\_planner}. The formation planner is ``blind", assigning ideal geometric locations without awareness of obstacles. The path planner is ``reactive", focused only on dodging immediate threats on its way to those locations. An attack drone (the red circle) positions itself at the target location assigned to a drone (F1). Drone F1's path planner detects the attacker and executes a drastic evasive maneuver due to strong repulsive forces. While locally successful in avoiding the attacker, this uncoordinated turn steers drone F1 directly into the flight path of another drone (F2), leading to a collision.

\vspace{-0.5cm}
\vspace{-0.5cm}
\section{Related Work}
Recent research on drone and swarm robotics testing has largely shifted from real-world environments to simulated ones, allowing for more efficient and cost-effective experimentation. \cite{Masamba2022Hybrid,Sun2023Test,Stolfi2023Optimising,Ma2022Decentralized,Na2022Bio-Inspired} For example, Na et al. \cite{Na2022Bio-Inspired} proposed a biologically-inspired collision avoidance strategy using deep reinforcement learning for self-driving vehicles, which, although effective in its domain, is limited to individual vehicle behaviors and does not scale to complex drone swarm systems. Zhang et al. \cite{zhang2023testing} introduced the ABLE method, which dynamically updates test objectives in autonomous driving. Still, its application is constrained to single vehicles and does not account for the interactions within a swarm.

In the domain of swarm fuzzing, tools like SWARMFLAWFINDER \cite{jung2022swarmflawfinder} apply fuzzing to swarm robotics by using DCC to guide attack strategies. However, SWARMFLAWFINDER suffers from high computational complexity, as it re-evaluates each scenario for every iteration, making it inefficient for large swarms or extended missions. Moreover, its reliance on only four fixed attack strategies and rigid parameters limits its adaptability in dynamic environments.

Fuzzing tools targeting individual drones \cite{Sheikhi2022Coverage-Guided,Schmidt2022StellaUAV:,Chambers2023HIFuzz:,Wang2023Toward, Khatiri2023Simulation-based,jung2022swarmflawfinder}, such as StellaUAV \cite{Schmidt2022StellaUAV:} and HiFuzz \cite{Chambers2023HIFuzz:}, primarily focus on vulnerabilities like buffer overflows and input validation errors in single drone systems. However, these tools are not suited for swarm systems, where inter-agent interactions and collective behaviors play a crucial role.

In the field of drone attack and defense, various methods have been proposed to address vulnerabilities at different levels, including network-level \cite{Fan2022Adversarial}, mission-level \cite{Islam2023A}, and device-level \cite{Duan2023In-Vehicle, Mykytyn2023GPS-Spoofing}. In this paper, we focus on the mission-level attack and defense mechanisms for drone swarms. RSFuzz addresses this by using swarm robustness to guide fuzz testing, improving the security and reliability of multi-drone systems and enhancing their resilience to external disruptions.
\section{Discussion}
The identification of key nodes requires constructing a constraint influence graph based on the swarm state, which improves failure detection but incurs additional computational overhead due to frequent graph updates. Each update necessitates recalculating the interactions among all drones in the swarm, which, as the swarm size increases, becomes increasingly expensive. While this is effective for small to medium-sized swarms, it may limit scalability for larger systems. In addition, both our framework and SWARMFLAWFINDER rely on gray-box fuzzing, which requires access to the algorithm's internal state to compute the swarm's next position, limiting applicability in black-box fuzzing.

	
\section{Conclusion}
In this paper, we propose a new fuzzing framework for multi-robot swarms, called RSFuzz, to detect logical vulnerabilities in swarm algorithms. We introduce the novel concept of swarm robustness based on behavioral constraints and use it to guide the generation and mutation of fuzzing scenarios. We evaluated RSFuzz on three open-source algorithms implemented in two different programming languages, and our experiments show that RSFuzz outperforms similar tools in terms of vulnerability detection success rate and efficiency. We also release the code and data for future research.
\section{Acknowledgement}
This work was supported by the National Key R\&D Program of China (Grant No. 2023YFB3107500), National Natural Science Foundation of China (Grant No. 92267204, Grant No.92467201), Key R\&D Program of Shandong Province (Grant No. 2024CXPT039), the Ministry of Education, Singapore under its Academic Research Fund Tier 2 (Award ID: T2EP20222-0037), National Natural Science Foundation of China (Youth Program) under Grant No. 62402367, and funded by China Scholarship Council.
\bibliographystyle{ieeetr}
\bibliography{references}

\vspace{12pt}

\end{document}